\documentclass[twocolumn, aps, prc, superscriptaddress, showpacs,
floatfix]{revtex4}

\usepackage{multirow}
\usepackage{epsfig}
\usepackage{amsmath}

\usepackage{subfigure}
\usepackage{latexsym}
\usepackage{feynmp}

\usepackage[normalem]{ulem}  
\usepackage[dvips]{color} 

\renewcommand\sout{\bgroup \color{red} \ULdepth=-.5ex \ULset}

\begin{document}

\title{Production of multi-charmed hadrons by recombination in heavy
ion collisions}

\author{Sungtae Cho}
\affiliation{Division of Science Education, Kangwon National
University, Chuncheon 24341, Korea}
\author{Su Houng Lee}
\affiliation{Department of Physics and Institute of Physics and
Applied Physics, Yonsei University, Seoul 03722, Korea}

\begin{abstract}

We study the production of multi-charmed hadrons by recombination
in heavy ion collisions by focusing on the production of
$\Xi_{cc}$, $\Xi_{cc}^*$, $\Omega_{scc}$, $\Omega_{scc}^*$,
$\Omega_{ccc}$ baryons and X(3872) mesons. Starting from the
estimation of yields for those hadrons at chemical freeze-out in
both the statistical and coalescence model, we evaluate their
transverse momentum distributions at mid-rapidity in the
coalescence model. We show that yields of multi-charmed hadrons in
heavy ion collisions at RHIC and LHC are large enough, and thereby
not only multi-charmed hadrons observed so far, e.g., the
$\Xi_{cc}$ but also those which have not been observed yet, can be
discovered sufficiently in heavy ion collisions. We also find that
the transverse momentum distribution ratio between various
multi-charmed hadrons sensitively reflects the interplay between
quark contents of corresponding hadrons as well as the transverse
momentum distribution of charm quarks at the hadronization point,
and therefore we insist that studying both the transverse momentum
distributions of multi-charmed hadrons themselves and transverse
momentum distribution ratios between various multi-charmed hadrons
provide us with useful information on hadron production mechanism
involving charm quarks in heavy ion collisions.
\end{abstract}

\pacs{25.75.-q, 14.65.Dw, 13.60.Rj}


\maketitle

\section{Introduction}

Relativistic heavy ion collision experiments have provided
exclusive opportunities to study a system of quantum chromodynamic
matter at high temperatures \cite{Adams:2005dq, Adcox:2004mh,
Gyulassy:2004zy}. Enormous energies available in heavy ion
collisions allow the system to reach a phase transition
temperature predicted by Lattice calculation \cite{Gupta:2011wh},
and as a result produce and sustain the so called quark-gluon
plasma composed of deconfined quarks and gluons for a certain time
during the experiments.

In order to investigate the properties of the quark-gluon plasma
many probes have been proposed, and among others a heavy quark
hadron has been considered to be one of the most useful probes in
understanding not only the quark-gluon plasma properties but also
the various aspects of phenomena in high energy heavy ion
collisions. The possible suppression of the $J/\psi$ production
caused by the color screening effects between charm and anti-charm
quarks in the quark-gluon plasma has been suggested as one of the
signatures for the formation of the system of free quarks and
gluons in heavy ion collision experiments \cite{Matsui:1986dk}.
More recently, members of charmonium states such as $\psi(2S)$ and
$\chi_c$ mesons are known to play a role as an indicator of the
temperature in the quark-gluon plasma phase since charmonium
states with different binding energies are expected to dissociate
at different temperatures in the system \cite{Satz:2005hx,
Karsch:2005nk, Mocsy:2007jz}.

As the energies achievable in relativistic heavy ion collisions
are increased, i.e., from $\sqrt{s_{NN}}=200$ GeV in Au-Au
collisions at the Relativistic Heavy Ion Collider (RHIC) to
$\sqrt{s_{NN}}=2.76$ TeV in Pb-Pb collisions at the Large Hadron
Collider (LHC), or from $\sqrt{s_{NN}}=2.76$ TeV to
$\sqrt{s_{NN}}=5.02$ TeV at LHC, heavy quarks are anticipated to
be more abundantly produced than ever before, and consequently the
possibilities of regenerating heavy quark hadrons from the
quark-gluon plasma are also expected to be increased
\cite{BraunMunzinger:2000px, Thews:2000rj, Andronic:2007bi}.
Therefore hadrons with heavy quarks become more important probes
in investigating the properties of the quark-gluon plasma, so that
it is necessary to understand in more detail the production of
heavy quark hadrons in high energy heavy ion collisions. In that
sense the coalescence model \cite{Greco:2003xt, Greco:2003mm,
Fries:2003vb, Fries:2003kq, Molnar:2003ff} which describes the
production of hadrons as the process of coalescing constituent
quarks in the quark-gluon plasma into hadrons using phase space
density functions has become a more relevant tool in investigating
the production of heavy quark hadrons.

In recent years the possibilities of observing various kinds of
exotic hadrons with heavy quarks, e.g., the $X(3872)$,
$Z_c(3900)$, $Z_c(4430)$, and $T_{cc}$ in heavy ion collision
experiments have been considered based on the coalescence model
\cite{Cho:2010db, Cho:2011ew, Cho:2017dcy}; it has been found that
the yield of those exotic hadrons are large enough to be observed
in heavy ion collisions, and the production yield of exotic
hadrons is strongly dependent on their structures. However, we
still find that production of even many normal hadrons with heavy
quarks have not been taken into account yet in heavy ion collision
experiments. Restricting our discussion to hadrons with charm
quarks, we have not observed multi-charmed hadrons such as
$\Xi_{cc}^*$, $\Omega_{scc}$, $\Omega_{scc}^*$, and $\Omega_{ccc}$
baryons even in elementary collisions though the existence of
these charmed baryons was predicted long time ago based on the
quark model with four flavors \cite{DeRujula:1975qlm}. In regard
to a doubly charmed baryon, the SELEX Collaboration reported the
first observation of the $\Xi_{cc}$ baryon in elementary
collisions in 2002 \cite{Mattson:2002vu}, which has not been
confirmed yet, while very recently the LHCb Collaboration has
reported the observation of the $\Xi_{cc}$ via the decay mode of
$\Lambda_c^+ K^-\pi^+\pi^+$ also in elementary collisions
\cite{Aaij:2017ueg}.

By this reason, it is necessary to study the production of normal
multi-charmed hadrons in relativistic heavy ion collisions, and
therefore we discuss in this work the production of the
$\Xi_{cc}$, $\Xi_{cc}^*$, $\Omega_{scc}$, $\Omega_{scc}^*$, and
$\Omega_{ccc}$ baryon in heavy ion collisions. The production of
the doubly charmed hadron, the $\Xi_{cc}$ has been considered
previously in heavy ion collisions based on the Schoedinger
equation \cite{Zhao:2016ccp}, and also based on the Boltzmann
transport equation \cite{Yao:2018zze}. Here, we focus on the
production of multi-charmed hadrons including also the $\Xi_{cc}$
by recombination in heavy ion collisions, and consider both their
yields and transverse momentum distributions.

We anticipate that when the mass of most doubly-charmed hadrons
are comparable to that of an exotic hadron, e.g., the $X(3872)$
meson, the yield of those hadron is also expected to be similar to
or smaller than the yield of the $X(3872)$ meson. The
$\Omega_{ccc}$ meson is an exception since it contains one more
charm quark, and therefore the yield of the $\Omega_{ccc}$ is
expected to be much smaller compared to that of the $X(3872)$
meson. Therefore, by comparing the yield of multi-charmed hadrons
to that of exotic hadrons obtained in Ref. \cite{Cho:2017dcy} we
can estimate the possibility of measuring those multi-charmed
hadrons in  heavy ion collisions.

Here when comparing multi-charmed hadrons to exotic hadrons we
restrict our discussion to the $X(3872)$ meson among many known
exotic hadrons. The $X(3872)$, first discovered by Belle
Collaboration in 2003 \cite{Choi:2003ue}, is one of exotic hadrons
whose structures have not been clearly understood. We still find
various possibilities for the structure of the $X(3872)$ meson; a
$\bar{D}^0D^{*0}$ hadronic molecule, a pure charmonium state, a
tetra-quark state, and a charmoniun-gluon hybrid state
\cite{Nielsen:2009uh}. In order to understand the structure of the
$X(3872)$ meson from the production in heavy ion collisions
various investigations have already been performed including those
involving the hadronic effects on the $X(3872)$ meson
\cite{Cho:2013rpa, Torres:2014fxa, Cleven:2019cre}. We also expect
to understand better the structure of the $X(3872)$ meson in this
work from studying the production of multi-charmed hadrons in
heavy ion collisions.

We also consider the transverse momentum distribution of the
$X(3872)$ meson as well as that of multi-charmed hadrons. Since it
has been found that the yield depends on the structure of hadrons,
it is expected that transverse momentum distributions also depends
on the structure of hadrons; we can obtain the yield after
integrating the transverse momentum distribution over all
transverse momenta. It has also been shown that transverse
momentum distributions of charmonium states are dependent on their
wave functions through the coalescence probability function, or
the Wigner function \cite{Cho:2014xha}. Therefore by evaluating
the transverse momentum distribution of multi-charmed hadrons we
expect to obtain useful information on those hadrons, e.g., the
dependence of the transverse momentum distribution on their
constituent heavy quarks. Especially for the $X(3872)$ meson we
consider two transverse momentum distributions of the $X(3872)$
assuming its structure to be either a four-quark state or a
two-quark state. As two different structures will lead to two
different transverse momentum distributions, future experimental
measurements on transverse momentum distribution of the $X(3872)$
are expected to discriminate the structure of the $X(3872)$.

Moreover, based on transverse momentum distributions of
multi-charmed hadrons we calculate the transverse momentum
distribution ratio between various multi-charmed hadrons, and also
the ratio between the $X(3872)$ and multi-charmed hadrons as has
been done previously between the anti-proton and pion. The ratio
between the anti-proton and pion is the baryon-to-meson ratio
whereas the ratio between the $X(3872)$ meson and multi-charmed
hadrons is the meson-to-baryon ratio. However, the ratio between
the $X(3872)$ and $\Xi_{cc}$ will retain the same quark contents
after cancelling the common quarks as in the ratio between the
anti-proton and pion thereby enabling us to investigate the
possibility of the enhanced production of the $X(3872)$ meson
compared to the $\Xi_{cc}$ baryon.

Since the quark structure of the $X(3872)$ meson in a four-quark
picture is considered to be $c\bar{c}q\bar{q}$, the $X(3872)$
meson is not a flavor exotic multi-charmed hadron, and therefore
it is more appropriate to consider an explicit multi-charmed
exotic hadron such as the $T_{cc}$ composed of $cc \bar{q}
\bar{q}$ instead of the $X(3872)$. However, the $T_{cc}$ has not
been observed experimentally yet, and moreover the yield and
transverse momentum distribution of the $X(3872)$ is almost same
as those of the $T_{cc}$ in the coalescence model under the
condition that transverse momentum distributions of charm and
anti-charm quarks are same. We only expect the slight difference
between the transverse momentum distribution of the $X(3872)$ and
that of the $T_{cc}$ at RHIC due to non-zero baryon chemical
potential. Therefore, we discuss in detail the yield and
transverse momentum distribution of the $T_{cc}$, but we focus
more on the $X(3872)$ rather than the $T_{cc}$, and adopt the
$X(3872)$ in comparing to normal multi-charmed hadrons.

The paper is organized as follows. In Sec. II, we first pay
attention to the yields, and estimate the values of multi-charmed
hadrons, the $\Xi_{cc}$, $\Xi_{cc}^*$, $\Omega_{scc}$,
$\Omega_{scc}^*$, $\Omega_{ccc}$ baryon and the X(3872) meson at
chemical freeze-out in both the statistical and coalescence model.
We discuss also the various yield ratio between multi-charmed
hadrons. In Sec. III, we focus on the transverse momentum
distribution in heavy ion collisions, and evaluate that of
multi-charmed hadrons mentioned above at mid-rapidity in the
coalescence model. Then, we obtain the transverse momentum
distribution ratios between multi-charmed hadrons, and investigate
the dependence of those ratios on the quark contents, numbers of
quarks, and so on in Sec. IV. Section V is devoted to conclusions.
We show the equivalence of the transverse momentum distribution of
a four-quark hadron on alternative relative coordinates in
Appendix A. In the paper we use sometimes the simplified notation
for the $X(3872)$ meson; $X_4$ for the $X(3872)$ meson in a
four-quark state, and $X_2$ for the $X(3872)$ meson in a two-quark
state.

\section{Production of multi-charmed hadrons from the quark-gluon plasma}

We evaluate yields of multi-charmed hadrons, $\Xi_{cc}$,
$\Xi_{cc}^*$, $\Omega_{scc}$, $\Omega_{scc}^*$, $\Omega_{ccc}$
baryons as well as $X(3872)$ and $T_{cc}$ mesons produced in
relativistic heavy ion collisions using both the statistical and
coalescence model in mid-rapidity. The statistical hadronization
model assuming hadron production in thermal and chemical
equilibrium at chemical freeze-out, has been very successful in
explaining the production yields of hadrons in heavy ion
collisions \cite{Andronic:2005yp}. In applying the statistical
hadronization model here for the estimation of the production
yields of multi-charmed hadrons we introduce additional charm
quark fugacities, $\gamma_c$ in order to take into account charm
quarks which are not in equilibrium in a quark-gluon plasma phase
due to their heavier masses compared to available temperatures in
a system. The yields are then given as
\begin{equation}
N^\mathrm{stat}_h = V_H \frac{g_h}{2 \pi^2} \int_0^\infty
\frac{p^2 dp}{\gamma_c^{-n}e^{E_h/T_H} \pm 1} \label{Nstat}
\end{equation}
where $g_h$ is the degeneracy factor of a hadron of species $h$,
$n$ number of charm quarks in the hadron, $V_H$ and $T_H$ are the
hadronization volume and temperature, respectively.
$E_h=\sqrt{m_h^2+p^2}$ in Eq. (\ref{Nstat}) is the energy of the
hadron of mass $m_h$. Here we consider the multi-charmed hadrons
produced at the hadronization temperature and volume $T_H$ = 162
MeV and $V_H$ = 2100 fm$^3$ at RHIC \cite{Andronic:2012dm} and
$T_H$ = 156 MeV and $V_H$ = 5380 fm$^3$ at LHC
\cite{Stachel:2013zma}, respectively. We assume that the total
numbers of charm quarks available from the initial hard collisions
are 4.1 at RHIC and 11 at LHC, which leads to the charm quark
fugacity factors $\gamma_c$ = 22 at RHIC and 39 at LHC
\cite{Cho:2017dcy}. All charm quarks produced at the initial hard
collisions are assumed to be conserved and fully distributed to
charmed hadrons including $D$, $D^*$, $D_s$ mesons, and
$\Lambda_c$ after chemical freeze-out \cite{Cho:2010db,
Cho:2011ew, Cho:2017dcy}.

We also consider yields of the $\Xi_{cc}$, $\Xi_{cc}^*$,
$\Omega_{scc}$, $\Omega_{scc}^*$, $\Omega_{ccc}$, $X(3872)$ and
$T_{cc}$ in the coalescence model which successfully explains the
enhanced production of the baryon compared to the meson in the
intermediate transverse momentum region \cite{Greco:2003xt,
Greco:2003mm, Fries:2003vb, Fries:2003kq}. Following Ref.
\cite{Cho:2017dcy} we assume that hadron productions by
coalescence occur at the critical temperature 166 MeV in the
volume 1790 (3530) fm$^3$ at RHIC (LHC) from quark constituents,
light quarks, strange quarks, and charm quarks of their masses 350
MeV, 500 MeV and 1500 MeV, respectively. We also adopt that the
number of light quarks available at hadronization is 302 (593),
that of strange quarks 176 (347) and that of charm quarks 4.1 (11)
at RHIC (LHC). Finally, by taking the charm quark oscillator
frequencies for the Wigner function, $\omega_c$ = 244 MeV for RHIC
and 278 MeV for LHC we evaluate the production yields of the
$\Xi_{cc}$, $\Omega_{scc}$, $\Omega_{ccc}$  baryons, $T_{cc}$ and
$X(3872)$ mesons, and show results in Table \ref{statcoalyields}.

\begin{table}[!h]
\caption{The $\Xi_{cc}$, $\Xi_{cc}^*$, $\Omega_{scc}$,
$\Omega_{scc}^*$, $\Omega_{ccc}$, $T_{cc}$ and $X(3872)$ yields at
mid-rapidity in both the statistical and coalescence model
expected at RHIC in $\sqrt{s_{NN}}=200$ GeV Au+Au collisions and
at LHC in $\sqrt{s_{NN}}=2.76$ TeV Pb+Pb collisions. }
\label{statcoalyields}
\begin{center}
\begin{tabular}{c|c|c|c|c}
\hline \hline
& \multicolumn{2}{|c}{RHIC} & \multicolumn{2}{|c}{LHC}   \\
\cline{2-5}
& Stat. & Coal. & Stat. & Coal. \\
\hline $\Xi_{cc}$ & $1.0\times 10^{-2}$ & $1.3\times 10^{-3}$
& $2.8\times 10^{-2}$ & $4.9\times 10^{-3}$  \\
$\Xi_{cc}^*$ & $6.4\times 10^{-3}$ & $9.0\times 10^{-4}$
& $1.8\times 10^{-2}$ & $3.3\times 10^{-3}$  \\
$\Omega_{scc}$ & $2.8\times 10^{-3}$ & $2.5\times 10^{-4}$
& $8.0\times 10^{-3}$ & $9.0\times 10^{-4}$  \\
$\Omega_{scc}^*$ & $1.5\times 10^{-3}$ & $1.6\times 10^{-4}$
& $4.3\times 10^{-3}$ & $6.0\times 10^{-4}$  \\
$\Omega_{ccc}$ & $1.1\times 10^{-4}$ & $1.1\times 10^{-6}$
& $4.0\times 10^{-4}$ & $5.3\times 10^{-6}$  \\
$T_{cc}$ & $8.9\times 10^{-4}$ & $5.3\times 10^{-5}$
& $2.7\times 10^{-3}$ & $1.3\times 10^{-4}$ \\
$X_2$ & $5.7\times 10^{-4}$ & $5.6\times 10^{-4}$
& $1.7\times 10^{-3}$ & $1.7\times 10^{-3}$ \\
$X_4$ & $5.7\times 10^{-4}$ & $5.3\times 10^{-5}$
& $1.7\times 10^{-3}$ & $1.3\times 10^{-4}$ \\
\hline \hline
\end{tabular}
\end{center}
\end{table}
We show two yields for the $X(3872)$ meson, one for the $X(3872)$
in a two-quark state, the $X_2$ and the other for the $X(3872)$
meson in a four-quark state, the $X_4$ \cite{Cho:2017dcy}. We
consider only a four-quark state for the $T_{cc}$ when evaluating
the yield of the $T_{cc}$ in the coalescence model.

In the statistical hadronization model 3621.4 MeV for the mass of
the $\Xi_{cc}$ \cite{Aaij:2017ueg}, 3648.0 MeV for the
$\Xi_{cc}^*$, 3679.0 MeV for the $\Omega_{scc}$, 3765.0 MeV for
the $\Omega_{scc}^*$, 4761.0 MeV for the $\Omega_{ccc}$,
\cite{Briceno:2012wt}, 3871.6 MeV for the $X(3872)$
\cite{Beringer:1900zz}, and 3797 MeV for the $T_{cc}$
\cite{Cho:2017dcy} are adopted. The mass of multi-charmed hadron
taken here is very close to that obtained in the recent analysis
\cite{Karliner:2018hos}. When evaluating yields of the $\Xi_{cc}$
and $\Omega_{scc}$, we have assumed the exclusive decay of a spin
3/2 baryon to a spin 1/2 baryon, similar to decay modes of $\Xi^*$
and $\Delta$ baryons \cite{Beringer:1900zz} as summarized in the
Table \ref{decaymodes}. We expect that the $\Omega_{scc}$ decays
to the baryon with one charm quark like the $\Xi_c$ without
decaying to the $\Xi_{cc}$. The $\Omega_{ccc}$ is expected to
decay to either the $\Xi_{cc}$ or the $\Omega_{scc}$, but the
yield of the $\Omega_{ccc}$ is much smaller compared to those of
the $\Xi_{cc}$ and $\Omega_{scc}$, and therefore we neglect the
contribution of the $\Omega_{ccc}$ decay to the yield of the
$\Xi_{cc}$ and $\Omega_{scc}$.

\begin{table}[!h]
\caption{Assumed decay modes of $\Xi_{cc}^*$ and $\Omega_{scc}^*$,
baryons similar to the known decay modes of $\Delta$ and $\Xi^*$
baryons \cite{Beringer:1900zz}. } \label{decaymodes}
\begin{center}
\begin{tabular}{c|c|c|c}
\hline \hline \multicolumn{2}{c}{Assumed decay modes} &
\multicolumn{2}{|c}
{Similar decay modes} \\
\hline $\Xi_{cc}^*\to\Xi_{cc}$ & $100\%$ & $\Delta\to N$
& $100\%$  \\
$\Omega_{scc}^*\to\Omega_{scc}$ & $100\%$ & $\Xi^*\to\Xi$ &
$100\%$ \\
\hline \hline
\end{tabular}
\end{center}
\end{table}
We see in Table \ref{statcoalyields} that the yield decreases with
increasing number of charm and light quarks in multi-charmed
hadrons in both the statistical and coalescence models, which is
attributable to the smaller probability to combine much heavier
and rarer charm quarks in the hadronization process; the yields of
the triply charmed hadron, the $\Omega_{ccc}$ baryon and the
four-quark hadron, the $X_4$ meson are smaller compared to those
of the $\Xi_{cc}$ and $\Xi_{cc}^*$ baryon. On the other hand, we
find that when the contribution from the $\Omega_{scc}^*$ decay is
not considered the yield of the $X_4$ is comparable to that of the
$\Omega_{scc}$ in the coalescence model, $8.2\times10^{-4}$ at
RHIC and $3.0\times10^{-4}$ at LHC; forming two more light quarks
of the constituent mass 350 MeV is similar to constructing one
more strange quark of the mass 500 MeV in addition to two charm
quarks. We also note that when the $X(3872)$ is considered to be a
normal meson composed of a charm and and an anti-charm quark
$X_2$, the yield in the coalescence model is almost same as that
in the statistical model.

As shown in Table \ref{statcoalyields}, yields in the quark
coalescence model are smaller than those in the statistical model,
reflecting the suppression effects in the quark coalescence
process \cite{Cho:2010db, Cho:2011ew, Cho:2017dcy}; the ratios of
the yield in the coalescence model compared to that in the
statistical model are 0.010 (0.013) for the $\Omega_{ccc}$, 0.089
(0.11) for the $\Omega_{scc}$, 0.13 (0.18) for the $\Xi_{cc}$,
0.060 (0.048) for the $T_{cc}$, and 0.093 (0.077) for the $X_4$ at
RHIC (LHC).

We find in Table \ref{statcoalyields} that the yield ratios
between the $X_4$ and the $\Xi_{cc}$ are 0.056 (0.061) in the
statistical model ,and 0.039 (0.027) in the coalescence model at
RHIC (LHC), and that between the $X_4$ and the $\Omega_{scc}$ are
0.20 (0.21) in the statistical model and 0.21 (0.15) in the
coalescence model while that between $\Omega_{ccc}$ and the
$\Xi_{cc}$ are 0.011 (0.014) in the statistical model and 0.00079
(0.0011) in the coalescence model at RHIC (LHC). We calculate the
yield ratios between multi-charmed hadrons introduced in Table
\ref{statcoalyields}, or the ratio between the yield of the
heavier hadron and that of the lighter one, and summarized in
Table \ref{yieldratios}. We see that yield ratios involving the
$\Omega_{ccc}$, or $\Omega_{ccc}/\Xi_{cc}$,
$\Omega_{ccc}/\Omega_{scc}$ and $\Omega_{ccc}/X_4$ at LHC are
always larger than those at RHIC in both the statistical and
coalescence model. Other ratios, except for the $\Omega_{ccc}$
ratios at RHIC, are comparable to or larger than those at LHC.

\begin{table}[!h]
\caption{The yield ratios between multi-charmed hadrons, the
$\Xi_{cc}$, $\Omega_{scc}$, $\Omega_{ccc}$ baryon, and the
$X(3872)$ meson at mid-rapidity in both the statistical and
coalescence model expected at RHIC in $\sqrt{s_{NN}}=200$ GeV
Au+Au collisions and at LHC in $\sqrt{s_{NN}}=2.76$ TeV Pb+Pb
collisions. } \label{yieldratios}
\begin{center}
\begin{tabular}{c|c|c|c|c|c|c}
\hline \hline Stat. & \multicolumn{2}{|c}{/$\Xi_{cc}$} &
\multicolumn{2}{|c}{/$\Omega_{scc}$} &
\multicolumn{2}{|c}{/$\Omega_{ccc}$} \\
\cline{2-7}
& RHIC & LHC & RHIC & LHC & RHIC & LHC  \\
\hline $\Omega_{scc}$ & 0.27 & 0.18 & & & & \\
$\Omega_{ccc}$ & 0.011 & 0.014 & 0.038 & 0.050 & & \\
$X_4$ & 0.056 & 0.039 & 0.20 & 0.21 & 5.3 & 4.3  \\
\hline \hline Coal. & \multicolumn{2}{|c}{/$\Xi_{cc}$} &
\multicolumn{2}{|c}{/$\Omega_{scc}$} &
\multicolumn{2}{|c|}{/$\Omega_{ccc}$} \\
\cline{2-7}
& RHIC & LHC & RHIC & LHC & RHIC & LHC \\
\hline $\Omega_{scc}$ & 0.29 & 0.18 & & & & \\
$\Omega_{ccc}$ & 0.00079 & 0.0011 & 0.0044 & 0.0059 & & \\
$X_4$ & 0.061 & 0.027 & 0.21 & 0.15 & 49 & 25 \\
\hline \hline
\end{tabular}
\end{center}
\end{table}

\section{Transverse momentum distributions of multi-charmed hadrons}

We consider in the coalescence model the transverse momentum
distribution of multi-charmed hadrons, $\Xi_{cc}$, $\Xi_{cc}^*$,
$\Omega_{scc}$, $\Omega_{scc}^*$, $\Omega_{ccc}$ baryons,
$X(3872)$ and $T_{cc}$ mesons produced from one or two light and
two or three charm quarks. Starting from the yield equation in the
coalescence model \cite{Greco:2003mm} we obtain the transverse
momentum distribution of the $\Xi_{cc}$, $\Xi_{cc}^*$,
$\Omega_{scc}$, $\Omega_{scc}^*$, $\Omega_{ccc}$ baryon, the
$X(3872)$ and $T_{cc}$ meson, respectively.

\subsection{Transverse momentum distributions of $\Xi_{cc}$,
$\Xi_{cc}^*$, $\Omega_{scc}$, $\Omega_{scc}^*$ and $\Omega_{ccc}$
baryons}

The yield for the $\Xi_{cc}$ baryon produced from one light quark
$l$ and two charm quarks, $c_1$ and $c_2$ is given by,
\begin{eqnarray}
&& N_{\Xi_{cc}}=g_{\Xi_{cc}}\int p_l\cdot d\sigma_l p_{c_1}\cdot
d\sigma_{c_1} p_{c_2}\cdot d\sigma_{c_2}\frac{d^3\vec
p_l}{(2\pi)^3E_l} \nonumber \\
&& \qquad\times \frac{d^3\vec p_{c_1}}{(2\pi)^3 E_{c_1}} \frac{d^3
\vec p_{c_2}}{(2\pi)^3 E_{c_2}} f_l(r_l,
p_l)f_{c_1}(r_{c_1}, p_{c_1}) \nonumber \\
&& \qquad\times f_{c_2}(r_{c_2}, p_{c_2})W_{\Xi_{cc}}(r_l,
r_{c_1}, r_{c_2}; p_l, p_{c_1}, p_{c_2}), \label{CoalGenXi}
\end{eqnarray}
with $d\sigma_q$ being the space-like hypersurface element for a
quark $q$. $f_q(r_q, p_q)$ is a covariant distribution function of
a quark $q$ satisfying the normalization condition $\int p_q\cdot
d\sigma_q d^3\vec p_q/((2\pi)^3E)f_q(r_q, p_q)=N_q$, the number of
quarks $q$ in the system. The factor $g_{\Xi_{cc}}$ takes into
account the possibility of forming the $\Xi_{cc}$ baryon from
constituent quarks, e.g., $g_{\Xi_{cc}}=2\times2/(2\cdot 3)^3$. In
the non-relativistic limit, Eq. (\ref{CoalGenXi}) is reduced to
\cite{Greco:2003mm, Greco:2003xt, Oh:2009zj}
\begin{eqnarray}
&& \frac{d^2N_{\Xi_{cc}}}{d^2\vec p_T}=\frac{g_{\Xi_{cc}}}{V^2}
\int d^3\vec r_1 d^3\vec r_2 d^2\vec p_{lT}d^2\vec p_{c_1 T}
d^2\vec p_{c_2 T} \nonumber \\
&& \qquad\quad \times \delta^{(2)}(\vec p_T-\vec p_{lT}-\vec
p_{c_1 T}-\vec p_{c_2 T})\frac{d^2N_l}{d^2 \vec p_{lT}} \nonumber \\
&& \qquad\quad \times\frac{d^2N_{c_1}}{d^2\vec p_{c_1 T}}
\frac{d^2N_{c_2}}{d^2\vec p_{c_2 T}}W_{\Xi_{cc}}(\vec r_1, \vec
r_2, \vec r_3, \vec k_1, \vec k_2, \vec k_3), \label{3CoalTrans}
\end{eqnarray}
with the assumption of the boost-invariant longitudinal momentum
distributions for quarks satisfying $\eta=y$, the Bjorken
correlation between spatial, $\eta$ and momentum $y$ rapidities.
In Eq. (\ref{3CoalTrans}) $\vec r_i$ and $\vec k_i$ are relative
distances and transverse momenta between quarks, respectively.
$\vec k_i$ is related to the transverse momenta of quarks through
the Lorentz transformation, from the transverse momenta in the
rest frame of the produced $\Xi_{cc}$ baryon, $\vec p_{iT}$ to
those in the fireball rest frame, $\vec p_{iT}'$
\cite{Scheibl:1998tk, Oh:2009zj}. We consider here the following
quark configuration which has been used in \cite{Cho:2010db,
Cho:2011ew},
\begin{eqnarray}
&& \vec R=\vec r_l+\vec r_{c_1}+\vec r_{c_2}, \nonumber \\
&& \vec r_1=\vec r_{c_1}-\vec r_{c_2}, \nonumber \\
&& \vec r_2=\frac{m_c\vec r_{c_1}+m_c\vec
r_{c_2}}{m_c+m_c}-\vec r_l, \nonumber \\
\label{rel_3coord}
\end{eqnarray}
for relative quark coordinates, and
\begin{eqnarray}
&& \vec k=\vec p_{lT}'+\vec p_{c_1 T}'+\vec p_{c_2 T}', \nonumber \\
&& \vec k_1=\frac{m_c\vec p_{c_1 T}'-m_c\vec p_{c_2 T}'}{m_c+m_c},
\nonumber \\
&& \vec k_2=\frac{m_l(\vec p_{c_1 T}'+\vec p_{c_2 T}')-(m_c+m_c)
\vec p_{lT}'}{m_l+m_c+m_c}, \label{rel_3moment}
\end{eqnarray}
for relative quark transverse momenta. Reduced masses
corresponding to the above configurations are
\begin{eqnarray}
&& \mu_1=\frac{m_cm_c}{m_c+m_c}, \quad \mu_2=
\frac{(m_c+m_c)m_l}{m_l+m_c+m_c}.
\end{eqnarray}
We adopt the following $s$-wave Wigner function constructed from
harmonic oscillator wave functions,
\begin{eqnarray}
&& W_{\Xi_{cc}}(\vec r_1, \vec r_2, \vec k_1, \vec k_2) \nonumber \\
&& \quad =8^2\exp{\bigg(-\frac{r_1^2}{\sigma_1^2}-\sigma_1^2k_1^2
\bigg)}\exp{\bigg(-\frac{r_2^2}{\sigma_2^2}-\sigma_2^2k_2^2\bigg)}
\label{wigner3}
\end{eqnarray}
with $\sigma_i^2=1/(\mu_i\omega)$, where $\omega$ is the oscillator
frequency of the harmonic oscillator wave function. After plugging
the Wigner function, Eq. (\ref{wigner3}) into Eq.
(\ref{3CoalTrans}), and carrying out the coordinate space
integration we obtain,
\begin{eqnarray}
&& \frac{d^2N_{\Xi_{cc}}}{d^2\vec p_T}=\frac{g_{\Xi_{cc}}}{V^2}
(2\sqrt{\pi})^6 (\sigma_1\sigma_2)^3\int d^2\vec p_{lT}d^2\vec
p_{c_1T}d^2\vec p_{c_2T} \nonumber \\
&& \qquad\quad \times \delta^{(2)}(\vec p_T-\vec p_{lT}-\vec
p_{c_1T}-\vec p_{c_2T})\frac{d^2N_l}{d^2 \vec
p_{lT}}  \nonumber \\
&& \qquad\quad \times\frac{d^2N_{c_1}}{d^2\vec p_{c_1T}}
\frac{d^2N_{c_2}}{d^2\vec p_{c_2T}}\exp{\bigg(-\sigma_1^2
k_1^2-\sigma_2^2k_2^2\bigg)}. \label{CoalTransXicc}
\end{eqnarray}
Similarly, the yield of the $\Xi_{cc}^*$ baryon produced from the
same quark constituents as the $\Xi_{cc}$ baryon is given by,
\begin{eqnarray}
&& \frac{d^2N_{\Xi_{cc}^*}}{d^2\vec
p_T}=\frac{g_{\Xi_{cc}^*}}{V^2} (2\sqrt{\pi})^6
(\sigma_1\sigma_2)^3\int d^2\vec p_{lT}d^2\vec
p_{c_1T}d^2\vec p_{c_2T} \nonumber \\
&& \qquad\quad \times \delta^{(2)}(\vec p_T-\vec p_{lT}-\vec
p_{c_1T}-\vec p_{c_2T})\frac{d^2N_l}{d^2 \vec
p_{lT}}  \nonumber \\
&& \qquad\quad \times\frac{d^2N_{c_1}}{d^2\vec p_{c_1T}}
\frac{d^2N_{c_2}}{d^2\vec p_{c_2T}}\exp{\bigg(-\sigma_1^2
k_1^2-\sigma_2^2k_2^2\bigg)}. \label{CoalTransXiccstar}
\end{eqnarray}
with $g_{\Xi_{cc}^*}=2\times4/(2\cdot 3)^3$ being the chance of
forming $\Xi_{cc}^*$ baryon from one light and two charm quarks.
The yield of the $\Omega_{scc}$ baryon produced from one strange,
$s$ and two charm quarks, $c_1$, $c_2$ is then given by,
\begin{eqnarray}
&& \frac{d^2N_{\Omega_{scc}}}{d^2\vec
p_T}=\frac{g_{\Omega_{scc}}}{V^2} (2\sqrt{\pi})^6
(\sigma_1\sigma_2)^3\int d^2\vec p_{c_1T}d^2\vec
p_{c_2T}d^2\vec p_{sT} \nonumber \\
&& \qquad\quad \times \delta^{(2)}(\vec p_T-\vec p_{c_1T}-\vec
p_{c_2T}-\vec p_{sT})\frac{d^2N_{c_1}}{d^2 \vec
p_{c_1T}}  \nonumber \\
&& \qquad\quad \times\frac{d^2N_{c_2}}{d^2\vec p_{c_2T}}
\frac{d^2N_s}{d^2\vec p_{sT}}\exp{\bigg(-\sigma_1^2
k_1^2-\sigma_2^2k_2^2\bigg)}, \label{CoalTransOmegascc}
\end{eqnarray}
with $g_{\Omega_{scc}}=2/(2\cdot 3)^3$ being the chance of forming
$\Omega_{scc}$ baryon from one strange and two charm quarks. Also
similarly, the yield of the $\Omega_{scc}^*$ baryon produced from
the same constituents as the $\Omega_{scc}$ is given by,
\begin{eqnarray}
&& \frac{d^2N_{\Omega_{scc}^*}}{d^2\vec
p_T}=\frac{g_{\Omega_{scc}^*}}{V^2} (2\sqrt{\pi})^6
(\sigma_1\sigma_2)^3\int d^2\vec p_{c_1T}d^2\vec
p_{c_2T}d^2\vec p_{sT} \nonumber \\
&& \qquad\quad \times \delta^{(2)}(\vec p_T-\vec p_{c_1T}-\vec
p_{c_2T}-\vec p_{sT})\frac{d^2N_{c_1}}{d^2 \vec
p_{c_1T}}  \nonumber \\
&& \qquad\quad \times\frac{d^2N_{c_2}}{d^2\vec p_{c_2T}}
\frac{d^2N_s}{d^2\vec p_{sT}}\exp{\bigg(-\sigma_1^2
k_1^2-\sigma_2^2k_2^2\bigg)}, \label{CoalTransOmegasccstar}
\end{eqnarray}
with the chance of forming $\Omega_{scc}^*$ baryon from one
strange and two charm quarks, $g_{\Omega_{scc}^*}=4/(2\cdot 3)^3$.
Finally the yield of the $\Omega_{ccc}$ baryon produced from three
charm quarks, $c_1$, $c_2$ and $c_3$ is,
\begin{eqnarray}
&& \frac{d^2N_{\Omega_{ccc}}}{d^2\vec
p_T}=\frac{g_{\Omega_{ccc}}}{V^2} (2\sqrt{\pi})^6
(\sigma_1\sigma_2)^3\int d^2\vec p_{c_1T}d^2\vec
p_{c_2T}d^2\vec p_{c_3T} \nonumber \\
&& \qquad\quad \times \delta^{(2)}(\vec p_T-\vec p_{c_1T}-\vec
p_{c_2T}-\vec p_{c_3T})\frac{d^2N_{c_1}}{d^2 \vec
p_{c_1T}}  \nonumber \\
&& \qquad\quad \times\frac{d^2N_{c_2}}{d^2\vec p_{c_2T}}
\frac{d^2N_{c_3}}{d^2\vec p_{c_3T}}\exp{\bigg(-\sigma_1^2
k_1^2-\sigma_2^2k_2^2\bigg)}, \label{CoalTransOmegaccc}
\end{eqnarray}
with $g_{\Omega_{ccc}}=4/(2\cdot 3)^3$ being the chance of forming
$\Omega_{ccc}$ baryon from three charm quarks.

\subsection{Transverse momentum distributions of the $X(3872)$
and $T_{cc}$ meson}

We also start with the yield equation in the coalescence model in
order to construct the transverse momentum distribution of the
$X(3872)$ meson produced from two light quarks $l$, $\bar{l}$ and
two charm quarks, $c$ and $\bar{c}$,
\begin{eqnarray}
&& N_X=g_X\int p_l\cdot d\sigma_l p_{\bar{l}}\cdot d\sigma_{
\bar{l}} p_c\cdot d\sigma_c p_{\bar{c}}\cdot d\sigma_{\bar{c}}
\nonumber \\
&& \qquad\times \frac{d^3\vec p_l}{(2\pi)^3E_l}\frac{d^3\vec
p_{\bar{l}}}{(2\pi)^3E_{\bar{l}}}\frac{d^3\vec p_c}{(2\pi)^3
E_c}\frac{d^3\vec p_{\bar{c}}}{(2\pi)^3E_{\bar{c}}} \nonumber \\
&& \qquad\times f_l(r_l, p_l)f_{\bar{l}}(r_{\bar{l}}, p_{\bar{l}})
f_c(r_c, p_c)f_{\bar{c}}(r_{\bar{c}}, p_{\bar{c}}) \nonumber \\
&& \qquad\times W_X(r_l, r_{\bar{l}}, r_c, r_{\bar{c}} ; p_l,
p_{\bar{l}}, p_c, p_{\bar{c}}), \label{CoalGenX}
\end{eqnarray}
with the same configurations introduced in Eq. (\ref{CoalGenXi});
$d\sigma_q$, $f_q(r_q, p_q)$ are, respectively the space-like
hypersurface element for a quark $q$, and a quark $q$ covariant
distribution function with the normalization $\int p_q\cdot
d\sigma_q d^3\vec p_q/((2\pi)^3E)f_q(r_q, p_q)=N_q$, the number of
$q$ quarks in the system. The factor $g_X$ covers the possibility
of forming the $X(3872)$ meson from constituent quarks, e.g.,
$g_X=3/(2\cdot 3)^4$. Eq. (\ref{CoalGenX}) is reduced in the
non-relativistic limit to \cite{Greco:2003mm, Greco:2003xt,
Oh:2009zj}
\begin{eqnarray}
&& \frac{d^2N_X}{d^2\vec p_T}=\frac{g_X}{V^3}\int d^3\vec r_1
d^3\vec r_2 d^3\vec r_3 d^2\vec p_{lT}d^2\vec p_{\bar{l}T}d^2\vec
p_{cT}d^2\vec p_{\bar{c}T} \nonumber \\
&& \qquad\quad \times \delta^{(2)}(\vec p_T-\vec p_{lT}-\vec
p_{\bar{l}T}-\vec p_{cT}-\vec p_{\bar{c}T})\frac{d^2N_l}{d^2 \vec
p_{lT}} \frac{d^2 N_{\bar{l}}}{d^2\vec p_{\bar{l}T}} \nonumber \\
&& \qquad\quad \times\frac{d^2N_c}{d^2\vec p_{cT}}
\frac{d^2N_{\bar{c}}}{d^2\vec p_{\bar{c}T}}W_X(\vec r_1, \vec r_2,
\vec r_3, \vec k_1, \vec k_2, \vec k_3), \label{4CoalTrans}
\end{eqnarray}
again with the boost-invariant longitudinal momentum distribution
assumption for quarks; $\eta=y$, the Bjorken correlation between
spatial, $\eta$ and momentum $y$ rapidities. In Eq.
(\ref{4CoalTrans}) relative transverse momenta, $\vec k_i$ are
related to the transverse momenta of quarks $\vec p_{iT}$ through
the Lorentz transformation from the transverse momenta in the rest
frame of the produced $X(3872)$ meson, $\vec p_{iT}$ to those in
the fireball rest frame, $\vec p_{iT}'$ \cite{Scheibl:1998tk,
Oh:2009zj}. The configuration of quarks inside the $X(3872)$ meson
is not unique, and possible quark arrangements for the $X(3872)$
meson are shown in the Appendix A. We adopt here the following
quark configuration for the $X(3872)$ meson,
\begin{eqnarray}
&& \vec R=\vec r_l+\vec r_{\bar{l}}+\vec r_c+\vec
r_{\bar{c}}, \nonumber \\
&& \vec r_1=\vec r_l-\vec r_{\bar{l}}, \nonumber \\
&& \vec r_2=\frac{m_l\vec r_l+m_{\bar{l}}\vec
r_{\bar{l}}}{m_l+m_{\bar{l}}}-\vec r_c, \nonumber \\
&& \vec r_3=\frac{m_l\vec r_l+m_{\bar{l}}\vec r_{\bar{l}}+m_c\vec
r_c}{m_l+m_{\bar{l}}+m_c}-\vec r_{\bar{c}}, \label{rel_coord}
\end{eqnarray}
for relative quark coordinates, and
\begin{eqnarray}
&& \vec k=\vec p_{lT}'+\vec p_{\bar{l}T}'+\vec p_{cT}'+\vec
p_{\bar{c}T}', \nonumber \\
&& \vec k_1=\frac{m_{\bar{l}}\vec p_{lT}'-m_l\vec
p_{\bar{l}T}'}{m_l+m_{\bar{l}}}, \nonumber \\
&& \vec k_2=\frac{m_c(\vec p_{lT}'+\vec
p_{\bar{l}T}')-(m_l+m_{\bar{l}})
\vec p_{cT}'}{m_l+m_{\bar{l}}+m_c}, \nonumber \\
&& \vec k_3=\frac{m_{\bar{c}}(\vec p_{lT}'+\vec p_{\bar{l}T}'+\vec
p_{cT}')-(m_l+m_{\bar{l}}+m_c)\vec
p_{\bar{c}T}'}{m_l+m_{\bar{l}}+m_c+m_{\bar{c}}},
\label{rel_moment}
\end{eqnarray}
for relative quark transverse momenta. Then, reduced masses
corresponding to the above configurations becomes,
\begin{eqnarray}
&& \mu_1=\frac{m_lm_{\bar{l}}}{m_l+m_{\bar{l}}}, \quad \mu_2=
\frac{(m_l+m_{\bar{l}})m_c}{m_l+m_{\bar{l}}+m_c}, \nonumber \\
&& \mu_3=\frac{(m_l+m_{\bar{l}}+m_c)m_{\bar{c}}}{m_l+m_{\bar{l}}+
m_c+m_{\bar{c}}}.
\end{eqnarray}
With the $s$-wave Wigner function made up of harmonic oscillator
wave functions,
\begin{eqnarray}
&& W_X(\vec r_1, \vec r_2, \vec r_3, \vec k_1, \vec k_2, \vec
k_3) \nonumber \\
&& \quad =8^3\exp{\bigg(-\frac{r_1^2}{\sigma_1^2}-\sigma_1^2k_1^2
\bigg)}\exp{\bigg(-\frac{r_2^2}{\sigma_2^2}-\sigma_2^2k_2^2\bigg)}
\nonumber \\
&& \quad\times\exp{\bigg(-\frac{r_3^2}{\sigma_3^2}-\sigma_3^2
k_3^2\bigg)} \label{wigner4}
\end{eqnarray}
we obtain the transverse momentum distribution of the $X(3872)$
meson,
\begin{eqnarray}
&& \frac{d^2N_X}{d^2\vec p_T}=\frac{g_X}{V^3}(2\sqrt{\pi})^9
(\sigma_1\sigma_2\sigma_3)^3\int d^2\vec p_{lT}d^2\vec
p_{\bar{l}T}d^2\vec p_{cT}d^2\vec p_{\bar{c}T} \nonumber \\
&& \qquad\quad \times \delta^{(2)}(\vec p_T-\vec p_{lT}-\vec
p_{\bar{l}T}-\vec p_{cT}-\vec p_{\bar{c}T})\frac{d^2N_l}{d^2 \vec
p_{lT}} \frac{d^2 N_{\bar{l}}}{d^2\vec p_{\bar{l}T}} \nonumber \\
&& \qquad\quad \times\frac{d^2N_c}{d^2\vec p_{cT}}
\frac{d^2N_{\bar{c}}}{d^2\vec p_{\bar{c}T}}\exp{\bigg(-\sigma_1^2
k_1^2-\sigma_2^2k_2^2-\sigma_3^2k_3^2\bigg)}.
\label{CoalTransX3872}
\end{eqnarray}
On the other hand, the transverse momentum distribution of the
$X(3872)$ meson produced from a charm and an anti-charm quark is
given by,
\begin{eqnarray}
&& \frac{d^2N_{X_2}}{d^2\vec p_T}=\frac{g_X}{V}(2\sqrt{\pi})^3
\sigma^3\int d^2\vec p_{cT}d^2\vec p_{\bar{c}T} \frac{2}{3}
\sigma^2k^2 e^{-\sigma^2 k^2} \nonumber \\
&& \qquad\quad \times\delta^{(2)}(\vec p_T-\vec p_{cT}-\vec
p_{\bar{c}T})\frac{d^2N_c}{d^2\vec p_{cT}}
\frac{d^2N_{\bar{c}}}{d^2\vec p_{\bar{c}T}}, \label{CoalTransX2}
\end{eqnarray}
which is same in form as the transverse momentum distribution of
the $p$-wave charmonium state $\chi_c$ \cite{Cho:2014xha}. In Eq.
(\ref{CoalTransX2}), $\vec k=(m_{\bar{c}}\vec p_{cT}'-m_c\vec
p_{\bar{c}T}')/(m_c+m_{\bar{c}})$ and
$\sigma^2=1/\mu\omega=(m_c+m_{\bar{c}})/m_c/m_{\bar{c}}/\omega$.
The transverse momentum distribution of the $T_{cc}$ is similar to
that of the $X(3872)$ composed of four quarks,
\begin{eqnarray}
&& \frac{d^2N_{T_{cc}}}{d^2\vec
p_T}=\frac{g_{T_{cc}}}{V^3}(2\sqrt{\pi})^9
(\sigma_1\sigma_2\sigma_3)^3\int d^2\vec p_{\bar{l}_1T}d^2\vec
p_{\bar{l}_2T}d^2\vec p_{c_1T} \nonumber \\
&& \qquad\quad \times d^2\vec p_{c_2T} \delta^{(2)}(\vec p_T-\vec
p_{\bar{l}_1T}-\vec p_{\bar{l}_2T}-\vec
p_{c_1T}-\vec p_{c_2T}) \nonumber \\
&& \qquad\quad \times\frac{d^2N_{\bar{l}_1}}{d^2 \vec
p_{\bar{l}_1T}} \frac{d^2N_{\bar{l}_2}}{d^2\vec p_{\bar{l}_2T}}
\frac{d^2N_{c_1}}{d^2\vec p_{c_1T}} \frac{d^2N_{c_2}}{d^2\vec
p_{c_2T}} \nonumber \\
&& \qquad\quad \times \exp{\bigg(-\sigma_1^2
k_1^2-\sigma_2^2k_2^2-\sigma_3^2k_3^2\bigg)}, \label{CoalTransTcc}
\end{eqnarray}
with $g_{T_{cc}}=3/(2\cdot 3)^4$ being the chance of forming the
$T_{cc}$ meson from two light and two charm quarks.

\subsection{Charm and light quark transverse momentum distributions}

\begin{figure}[!t]
\begin{center}
 \includegraphics[width=0.50\textwidth]{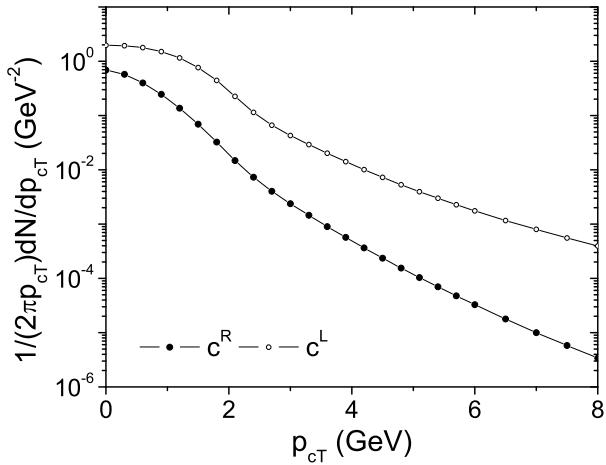}
\end{center}
\caption{(a) Transverse momentum distributions of charm quarks
$1/(2\pi p_{cT})dN/dp_{cT}$ at RHIC and LHC.} \label{dN2dpcT2}
\end{figure}

In order to evaluate the transverse momentum distribution of a
multi-charmed hadron we need the information on the transverse
momentum distribution of both light and charm quarks in the
system. We introduce here the following transverse momentum
distributions at mid-rapidities \cite{Plumari:2017ntm},

\begin{widetext}
\begin{eqnarray}
&& \frac{d^2N_c^R}{d^2 \vec p_{cT}}=\left\{
\begin{array}{ll}
0.69e^{(-1.22p_{cT}^{1.57})} & \quad p_{cT} \le 1.85~\textrm{GeV} \\
1.08e^{(-3.04p_{cT}^{0.71})}+3.79(1.0+p_{cT}^{2.02})^{-3.48} &
\quad p_{cT} > 1.85~\textrm{GeV} \\
\end{array} \right. \nonumber \\
&& \frac{d^2N_c^L}{d^2 \vec p_{cT}}=\left\{
\begin{array}{ll}
1.97e^{(-0.35p_{cT}^{2.47})} & ~p_{cT} \le 1.85~\textrm{GeV} \\
7.95e^{(-3.49p_{cT}^{3.59})}+87335(1.0+p_{cT}^{0.5})^{-14.31} &
~p_{cT} > 1.85~\textrm{GeV} \\
\end{array} \right. \label{dNcdpT}
\end{eqnarray}
\end{widetext}
for charm quarks shown in Fig. \ref{dN2dpcT2}. $d^2N_q/d^2\vec
p_{qT}$ with a superscript $R$ and $L$ in Eq. (\ref{dNcdpT})
represents a charm quark transverse momentum distribution at RHIC
and LHC, respectively.

For the light quark transverse momentum distribution we adopt the
following thermal distribution,
\begin{equation}
\frac{d^2N_l}{d^2 p_T}=g_l\frac{V}{(2\pi)^3}m_Te^{-m_T/T_{eff}},
\label{dNldpT}
\end{equation}
with the color-spin degeneracy of light quarks, $g_l=6$ and the
transverse mass $m_T=\sqrt{p_T^2+m^2}$. We use the same
coalescence volume 1790 and 3530 $\rm{fm}^3$ for RHIC and LHC,
respectively introduced in Ref. \cite{Cho:2017dcy}. $T_{eff}$ in
Eq. (\ref{dNldpT}) is an effective temperature which we take here
as $T_{eff}=177$ MeV for both RHIC and LHC in order to take into
account collective flow effects of the quark-gluon plasma. Even
though the same effective temperature is applied for both RHIC and
LHC, the larger flow effects at LHC have been taken into account
since freeze-out temperatures estimated in the statistical
hadronization model are 162 and 156 MeV at RHIC and LHC.

Light quarks also have a power law type transverse momentum
distribution at high transverse momentum. However, for such
effects to be visible in a charmed hadron, the momentum of a charm
quark or a charmed hadron itself should have an even larger
transverse momentum because the transverse momentum of heavy quark
hadrons is mostly dominated by the momentum of the heavy quarks.
Therefore, it is reasonable to consider only light quarks with an
exponential transverse momentum distribution when we are
interested in the transverse momentum distribution of a heavy
quark hadron up to about 8 GeV.

Using the above transverse momentum distributions with the light
quark mass 300 MeV, we obtain the total number of charm quarks
available in the system, $N_c=2.00$ at RHIC and $N_c=14.9$ at LHC,
and also the total number of light quarks $N_l=298$ at RHIC and
$N_{l}=588$ at LHC comparable to those in Ref. \cite{Cho:2017dcy}.

Finally, we have to determine the oscillator frequency,
$\omega_c$. Since the oscillator frequency is related to the size
of the hadron in the coalescence model, it is mostly determined
from the relation between the mean square distance $\langle
r^2\rangle$ and $\sigma$; for a $D$ meson $\langle
r^2\rangle=3/2\sigma^2= 3/2/\mu/\omega_c$. However, we take here
the oscillator frequency which enables all charm quarks at zero
transverse momentum to get hadronized entirely by quark
coalescence \cite{Oh:2009zj, Plumari:2017ntm}. It has been found
that hadrons are produced via two different hadron production
mechanisms in heavy ion collisions, one by quark coalescence or
the other by fragmentation \cite{Greco:2003mm, Greco:2003xt,
Fries:2003kq, Fries:2003vb}. Since the hadron production by quark
coalescence is dominant at low transverse momenta, it is natural
to expect exclusive hadron production at zero transverse momentum
by quark coalescence.

In extracting those oscillator frequencies we consider the
transverse momentum distributions for four open charm mesons, $D$,
$D^*$, $D_s$, and $D_s^*$, and ten charm baryons, $\Lambda_c$,
$\Sigma_c(2455)$, $\Sigma_c(2520)$, $\Lambda_c(2595)$,
$\Lambda_c(2625)$, $\Xi_c$, $\Xi_c'$, $\Xi_c(2645)$, $\Omega_c$,
and $\Omega_c(2770)$. Charmonium states and multi-charmed hadrons
are not taken into account in the evaluation of the oscillator
frequency since the transverse momentum distributions of charmonium
states or multi-charmed hadrons are negligible compared to those
of charmed hadrons mentioned above. We find that oscillator
frequencies, $\omega_c=0.078$ GeV at RHIC and $\omega_c=0.076$ GeV
at LHC guarantee the consumption of all charm quarks at zero
transverse momentum entirely by quark coalescence.

The oscillator frequencies $\omega_c=0.078$ or 0.076 GeV are
smaller than those in Ref. \cite{Cho:2017dcy}, $\omega_c=0.244$ or
0.276 GeV obtained on two conditions; the sum of yields for
charmed hadrons, e.g., $D$, $D^*$, and $D_s$ mesons and
$\Lambda_c$ baryons after hadronization in the statistical
hadronization model is equal to the total number of charm quarks
available in the quark-gluon plasma before hadronization, and the
yield of $\Lambda_c$ including the feed-down contributions in the
coalescence model agrees with that in the statistical model. As
has already been pointed out in Refs. \cite{Oh:2009zj,
Plumari:2017ntm} smaller oscillator frequencies obtained on the
requirement that all charm quarks at zero transverse momentum
should be used up entirely by quark coalescence give the size of
charmed hadrons relatively larger than the real size of those
hadrons. When the charm quark oscillator frequencies
$\omega_c=0.078$ or 0.076 GeV are applied, the size of the $D$
meson, $\sqrt{\langle r^2\rangle}=1.73$ fm, also larger than the
assumed size of $D$ mesons, $\sim$1.0 fm.

\begin{figure}[!t]
\begin{center}
\includegraphics[width=0.50\textwidth]{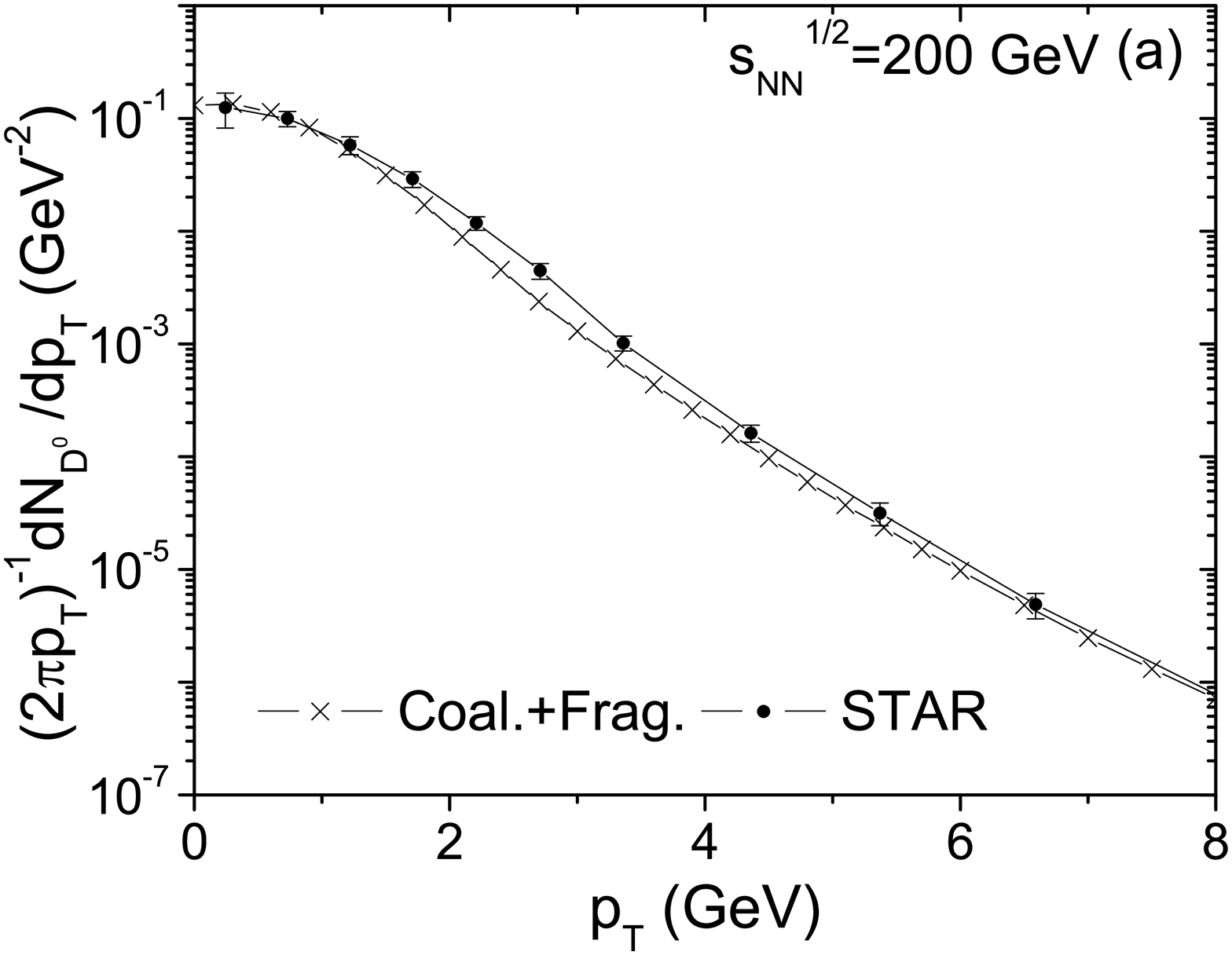}
\includegraphics[width=0.55\textwidth]{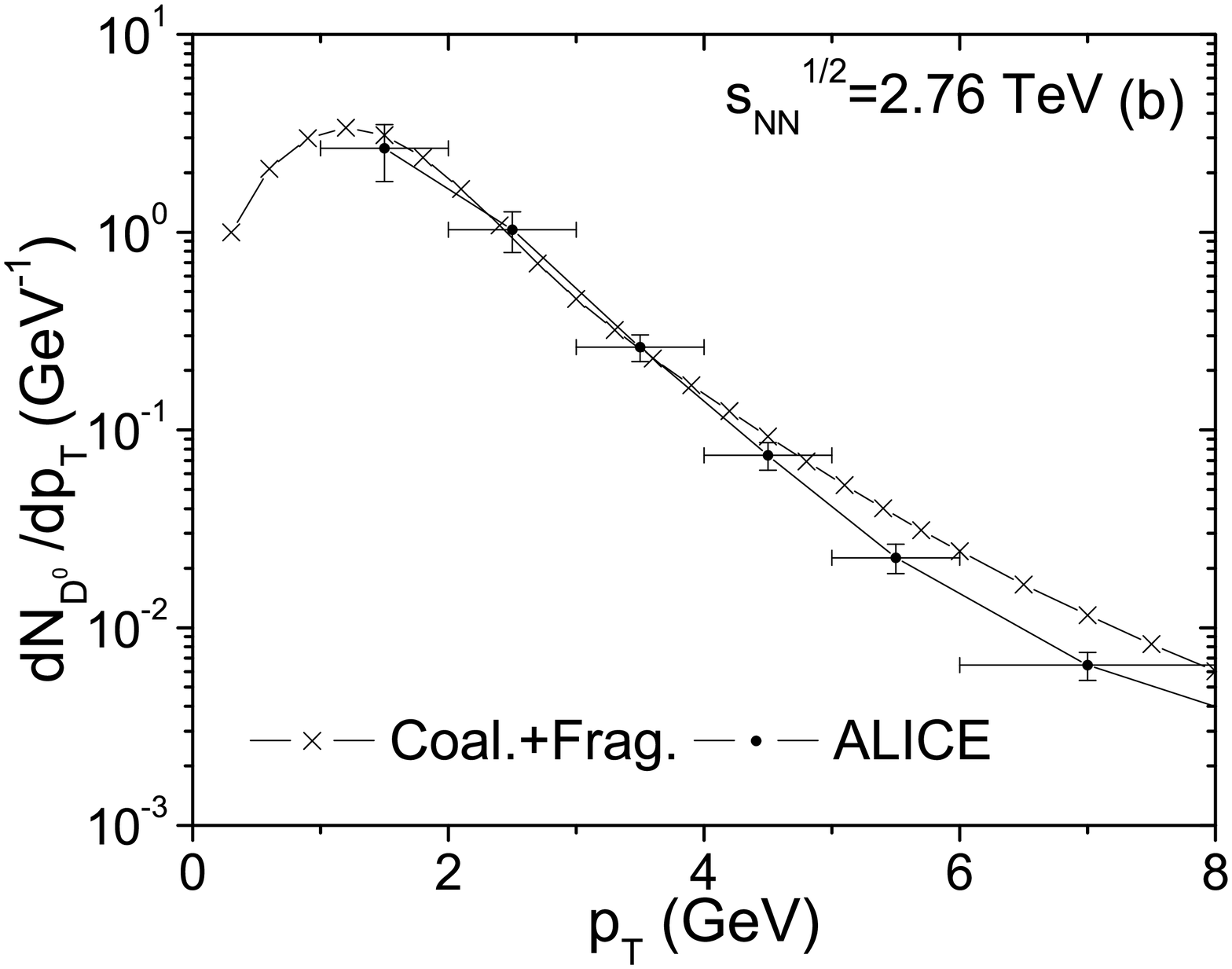}
\end{center}
\caption{(a) Transverse momentum distributions of $D^0$ mesons,
$(2\pi p_T)^{-1}dN_D/dp_T$ at RHIC and (b) those multiplied by
$2\pi p_T$, $dN_D/dp_T$ at LHC.} \label{pTdistribution_D0}
\end{figure}

With these charm quark oscillator frequencies we evaluate the
transverse momentum distributions of $D^0$ mesons, and compare
those with experimental measurements at RHIC \cite{Adam:2018inb}
and LHC \cite{Adam:2015sza}. We take into account here the
feed-down contribution from $D^*$ mesons as well as that from the
production of $D$ mesons by fragmentation. We show results in Fig.
\ref{pTdistribution_D0}.

As we see in Fig. \ref{pTdistribution_D0}, transverse momentum
distributions of $D$ mesons evaluated here in the coalescence
model based on the charm and light quark transverse momentum
distributions, Eqs. (\ref{dNcdpT}) and (\ref{dNldpT}) with charm
quark oscillator frequencies $\omega_c=0.078$ or 0.076 GeV agree
reasonably well with those measurements at RHIC and LHC. The
transverse momentum distribution of $D^0$ mesons at RHIC is
slightly smaller than the experimental measurement at intermediate
transverse momenta whereas that at LHC deviates from measurements
within errors at high transverse momenta.

\subsection{Transverse momentum distributions of multi-charmed hadrons}

Now we evaluate transverse momentum distributions of $\Xi_{cc}$,
$\Xi_{cc}^*$, $\Omega_{scc}$, $\Omega_{scc}^*$, $\Omega_{ccc}$
baryons, $X(3872)$, and $T_{cc}$ mesons produced by quark
recombination using Eqs. (\ref{CoalTransXicc}),
(\ref{CoalTransXiccstar}), (\ref{CoalTransOmegascc}),
(\ref{CoalTransOmegasccstar}), (\ref{CoalTransOmegaccc}),
(\ref{CoalTransX3872}), (\ref{CoalTransX2}) and
(\ref{CoalTransTcc}). In the calculation we use light quark mass
$m_l=m_{\bar{l}}=300$ MeV, charm quark mass $m_c=m_{\bar{c}}=1500$
MeV, and the volume $1790$ fm$^3$ for RHIC and $3530$ fm$^3$ for
LHC.
\begin{figure}[!]
\begin{center}
\includegraphics[width=0.51\textwidth]{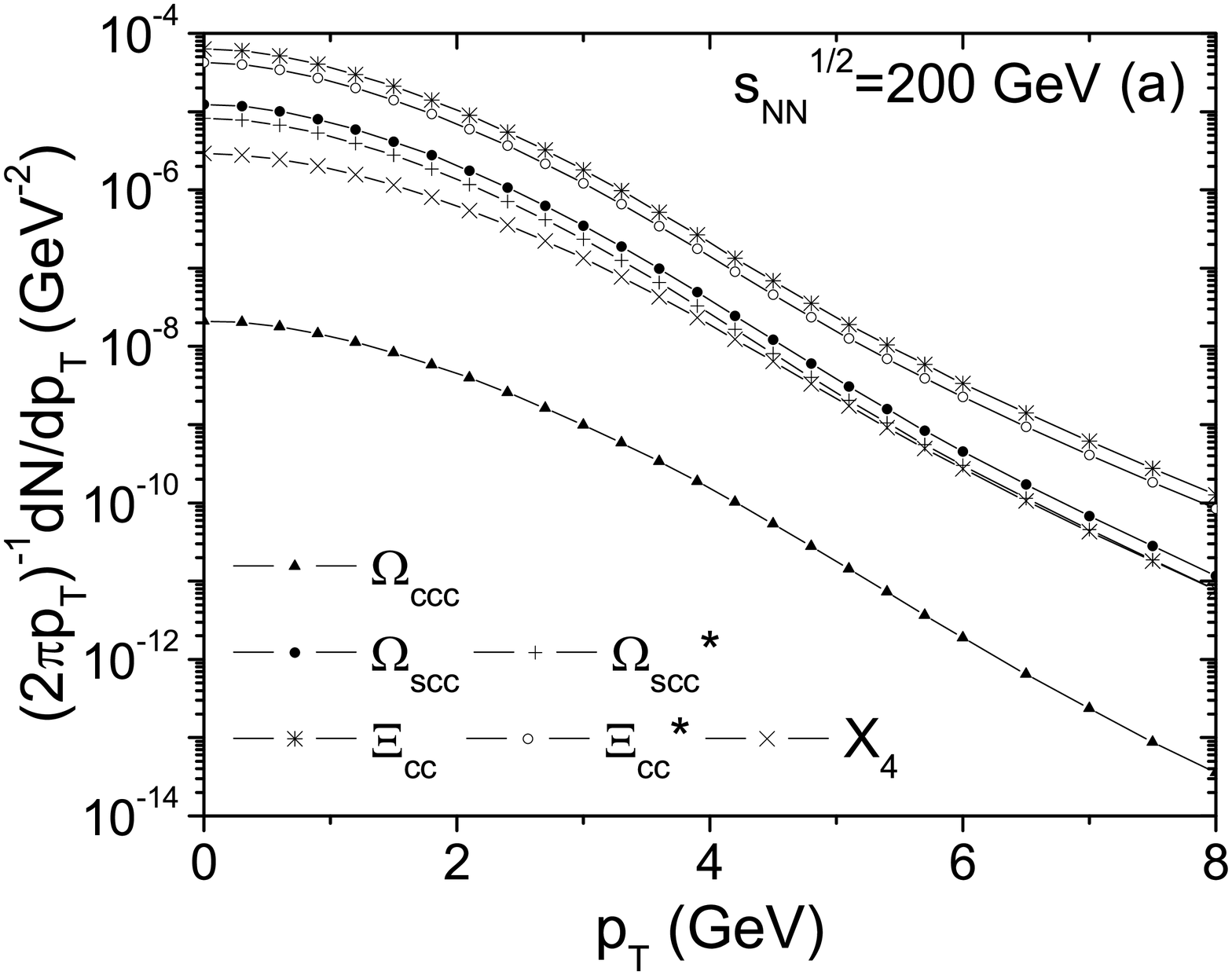}
\includegraphics[width=0.51\textwidth]{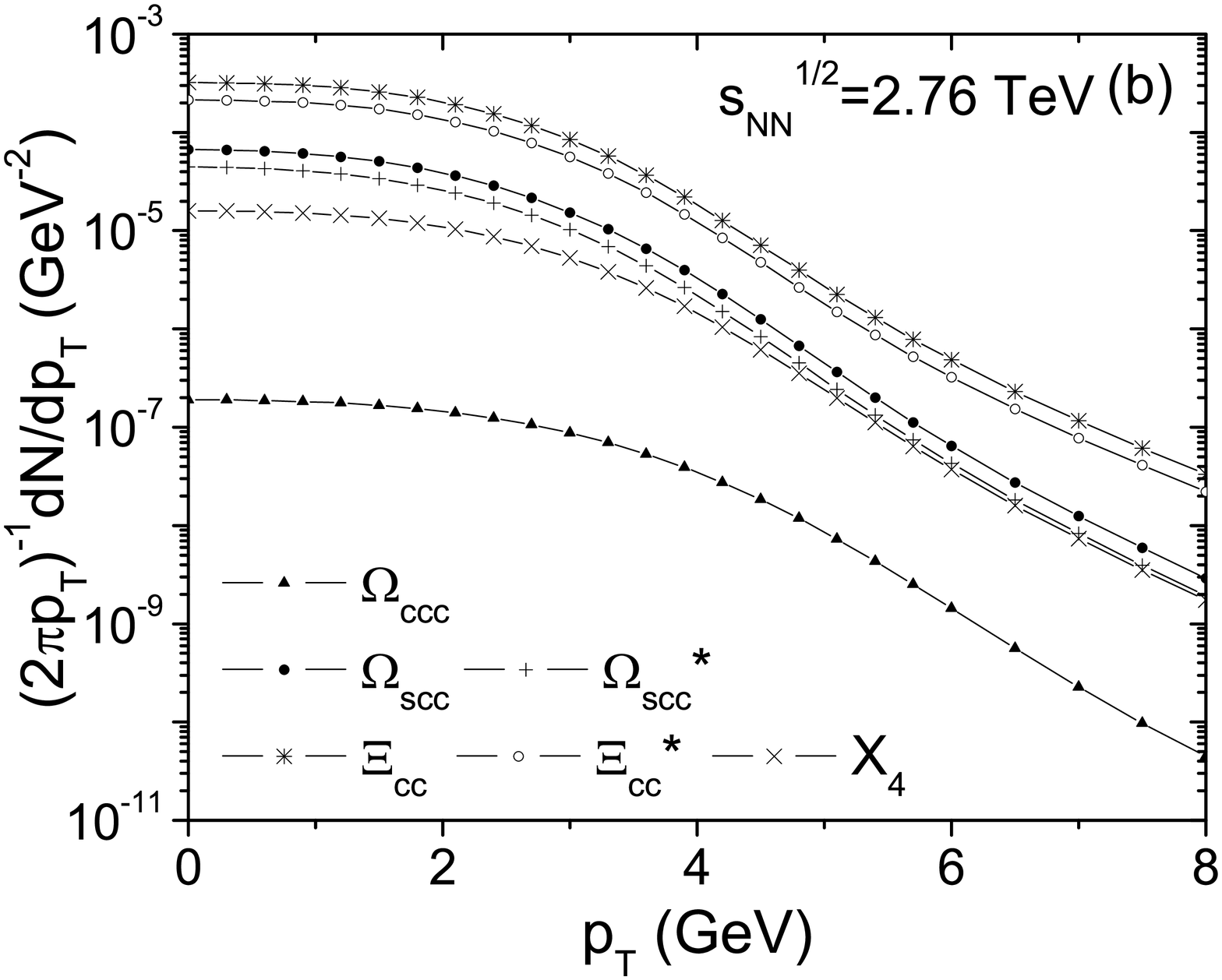}
\end{center}
\caption{Transverse momentum distributions, $(2\pi
p_T)^{-1}dN/dp_T$ of multi-charmed hadrons, $\Xi_{cc}$,
$\Xi_{cc}^*$, $\Omega_{scc}$, $\Omega_{scc}^*$, $\Omega_{ccc}$
baryons, and a $X(3872)$ meson in a four-quark state, $X_4$ at
RHIC (a) and at LHC (b). } \label{MultiCharm_pT}
\end{figure}

We show in Fig. \ref{MultiCharm_pT} transverse momentum
distributions, $(2\pi p_T)^{-1}dN/dp_T$ of multi-charmed hadrons,
$\Xi_{cc}$, $\Xi_{cc}^*$, $\Omega_{scc}$, $\Omega_{scc}^*$,
$\Omega_{ccc}$ baryons, and the $X(3872)$ meson in a four-quark
state, $X_4$ at both RHIC $\sqrt{s_{NN}}$=200 GeV and LHC
$\sqrt{s_{NN}}$=2.76 TeV. Here we have taken into account
feed-down contributions for transverse momentum distributions of
the $\Xi_{cc}$ and $\Omega_{scc}$ baryon from their spin 3/2
hadrons, $\Xi_{cc}^*$ and $\Omega_{scc}^*$ baryons, respectively.
However, we assume that transverse momentum distributions of the
daughter hadrons, $\Xi_{cc}$ and $\Omega_{scc}$ baryons are almost
same as those of $\Xi_{cc}^*$ and $\Omega_{scc}^*$ baryons after
the $\Xi_{cc}^*$ and $\Omega_{scc}^*$ baryon decays to the
$\Xi_{cc}$ and $\Omega_{scc}$ baryon, respectively since
$\Xi_{cc}$ and $\Omega_{scc}$ baryons are much heavier than the
other daughter hadron in the decay process.

As we see in Fig. \ref{MultiCharm_pT}, all transverse momentum
distributions of charmed hadrons at LHC are larger than those at
RHIC due to the larger number of charm and light quarks available
at LHC compared to that at RHIC. It has been found that the yield
of a hadron with more quarks is suppressed since the probability
to combine more quarks to form a multi-quark hadron decreases as
the number of quarks within a hadron is increased
\cite{Cho:2010db, Cho:2011ew, Cho:2017dcy}. Therefore it is
expected that the transverse momentum distribution of normal
hadrons composed of three quarks, $\Xi_{cc}$, $\Xi_{cc}^*$,
$\Omega_{scc}$, and $\Omega_{scc}^*$ baryons is larger than that
of four-quark hadrons, $X(3872)$ mesons at both RHIC and LHC. We
find that transverse momentum distributions of the $\Xi_{cc}$ and
$\Xi_{cc}^*$ baryon is larger than that of the $X_4$ meson by an
order of magnitude at both RHIC and LHC as expected but the
transverse momentum distribution of the $X_4$ meson is larger than
that of the $\Omega_{ccc}$ baryon by two orders of magnitude at
both RHIC and LHC. The effect from the much smaller abundance of
charm quarks in the system compared to that of light quarks by a
factor of hundreds at RHIC and LHC overwhelms the meaningful
contribution from the larger possibility for forming a hadron
composed of three charm quarks compared to the relatively smaller
possibility for forming a hadron with four quarks, leading to the
smaller transverse momentum distribution of the $\Omega_{ccc}$
baryon compared to that of the $X(3872)$ meson.

\begin{figure}[!t]
\begin{center}
\includegraphics[width=0.51\textwidth]{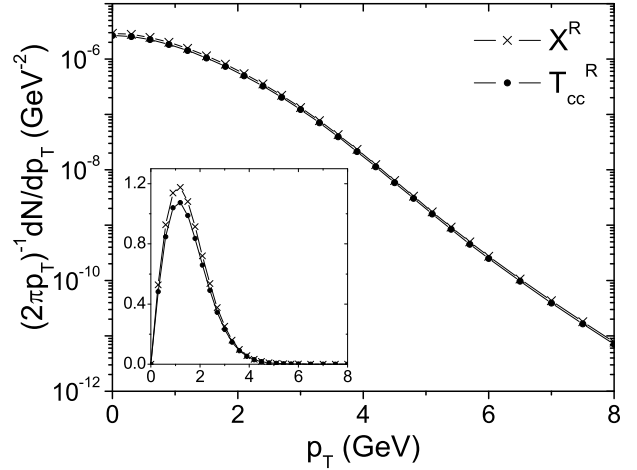}
\end{center}
\caption{Transverse momentum distributions of the $X(3872)$ in a
four-quark state, $X_4$ and the $T_{cc}$ meson for
$\sqrt{s_{NN}}$=200 GeV at RHIC. In the inset $dN/dp_T$ of the
$X_4$ and $T_{cc}$ in unit of $10^{-5}$GeV$^{-1}$ are shown.}
\label{XandTcc}
\end{figure}
We have also obtained the transverse momentum distribution of the
$T_{cc}$ meson, similar to that of the four-quark $X(3872)$ meson.
Since the same transverse momentum distributions of charm quarks
at RHIC and LHC, Eq. (\ref{dNcdpT}) are introduced, the difference
of the transverse momentum distribution between the $T_{cc}$ and
the $X_4$ is originated from the light quark/anti-quark
distributions, or the non-zero baryon chemical potential. We show
in Fig. \ref{XandTcc} the transverse momentum distributions of
$X_4$ and $T_{cc}$ mesons at RHIC. As expected the transverse
momentum distribution of the $T_{cc}$ is slightly smaller than
that of the $X_4$ due to the smaller number of anti-light quarks
caused by the baryon chemical potential, $\mu_l=24$ MeV
\cite{Andronic:2012dm}. We do not show the transverse momentum
distribution of the $T_{cc}$ at LHC since there exists no
difference in the transverse momentum distributions between $X_4$
and $T_{cc}$ mesons with the zero baryon chemical potential.

\begin{table}[!h]
\caption{Total yields of the $\Xi_{cc}$, $\Xi_{cc}^*$,
$\Omega_{scc}$, $\Omega_{scc}^*$, $\Omega_{ccc}$ baryon, the
$T_{cc}$ and $X(3872)$ meson at mid-rapidity obtained by
integrating the transverse momentum distributions shown in Fig.
\ref{MultiCharm_pT} over all transverse momenta at RHIC in
$\sqrt{s_{NN}}=200$ GeV Au+Au collisions and at LHC in
$\sqrt{s_{NN}}=2.76$ TeV Pb+Pb collisions. }
\label{charmedpTyields}
\begin{center}
\begin{tabular}{c|c|c}
\hline \hline
& RHIC & LHC   \\
\hline $\Xi_{cc}$ & $4.4\times 10^{-4}$ & $6.7\times 10^{-3}$   \\
$\Xi_{cc}^*$ & $2.9\times 10^{-4}$ & $4.5\times 10^{-3}$   \\
$\Omega_{scc}$ & $8.6\times 10^{-5}$ & $1.3\times 10^{-3}$   \\
$\Omega_{scc}^*$ & $5.7\times 10^{-5}$ & $8.5\times 10^{-4}$   \\
$\Omega_{ccc}$ & $1.7\times 10^{-7}$ & $5.9\times 10^{-6}$   \\
$T_{cc}$ & $2.2\times 10^{-5}$ & $3.8\times 10^{-4}$  \\
$X_4$ & $2.4\times 10^{-5}$ & $3.8\times 10^{-4}$  \\
$X_2$ & $2.6\times 10^{-4}$ & $4.5\times 10^{-3}$  \\

\hline \hline
\end{tabular}
\end{center}
\end{table}

We calculate the total yield of each charmed hadron by integrating
out transverse momentum distributions shown in Fig.
\ref{MultiCharm_pT} over all transverse momenta, and summarize the
results in Table \ref{charmedpTyields}. Here we have also taken
into account feed-down contributions for the $\Xi_{cc}$ and
$\Omega_{scc}$ baryon from their spin 3/2 hadrons, $\Xi_{cc}^*$
and $\Omega_{scc}^*$ baryons, respectively. As we see in Table
\ref{charmedpTyields}, the yields of the four-quark $X(3872)$ and
the $T_{cc}$ at RHIC are now slightly different. We note that the
yield of the $\Omega_{scc}$ is slightly larger than that of the
$X_4$. The yields based on the transverse momentum distributions
are smaller and larger compared to those yields in Table
\ref{statcoalyields} at RHIC and LHC, respectively, and this
difference must be mostly due to different numbers of charm quarks
introduced in the system, 2.0 vs. 4.1 at RHIC and 14.9 vs. 11.0 at
LHC.

\section{Transverse momentum distribution ratios}

\subsection{Transverse momentum distribution ratios between
multi-charmed hadrons}

Based on transverse momentum distributions of multi-charmed
hadrons as shown in Fig. \ref{MultiCharm_pT} we evaluate various
transverse momentum distribution ratios between multi-charmed
hadrons, similar to the anti-proton/pion ratio which shows an
increase up to unity in the intermediate transverse momenta due to
the contribution from two hadron production processes, the
recombination and fragmentation \cite{Greco:2003xt, Greco:2003mm,
Fries:2003kq, Fries:2003vb}. The transverse momentum distribution
ratio between an anti-proton and a pion is the ratio between a
three-light quark and a two-light quark hadrons,
$\bar{q}\bar{q}\bar{q}/q\bar{q}$ whereas the ratio evaluated here
is, for example, the ratio between a two-charm and two-light quark
and a two-charm and one-light quark hadrons,
$cc\bar{q}\bar{q}/ccq$, or $c\bar{c}q\bar{q}/ccq$,
$X(3872)/\Xi_{cc}$ and so on. After dividing out the same flavors
appearing in the numerator and denominator, since both ratios
simply retain the same kind of quark ratios, it is anticipated
that two ratios have similar behavior in the intermediate
transverse momentum region. We expect to obtain useful information
on constituent quarks in different hadrons by comparing the ratio
of normal hadrons produced by recombination in heavy ion
collisions, the baryon/meson ratio, to the ratio between the
four-quark $X(3872)$ meson and the $\Xi_{cc}$ baryon, the
meson/baryon ratio.

\begin{widetext}

\begin{figure}[!t]
\begin{center}
\includegraphics[width=0.495\textwidth]{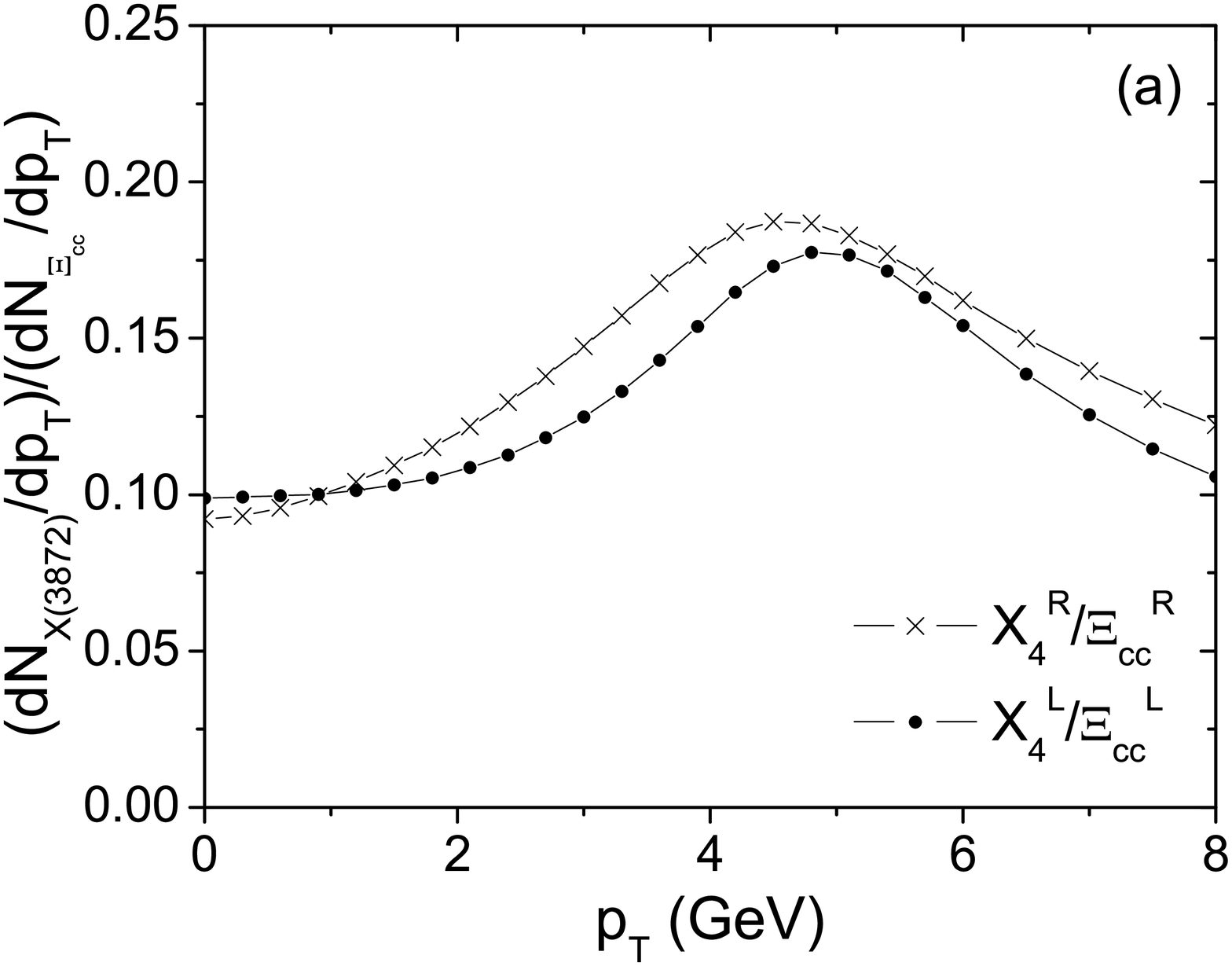}
\includegraphics[width=0.495\textwidth]{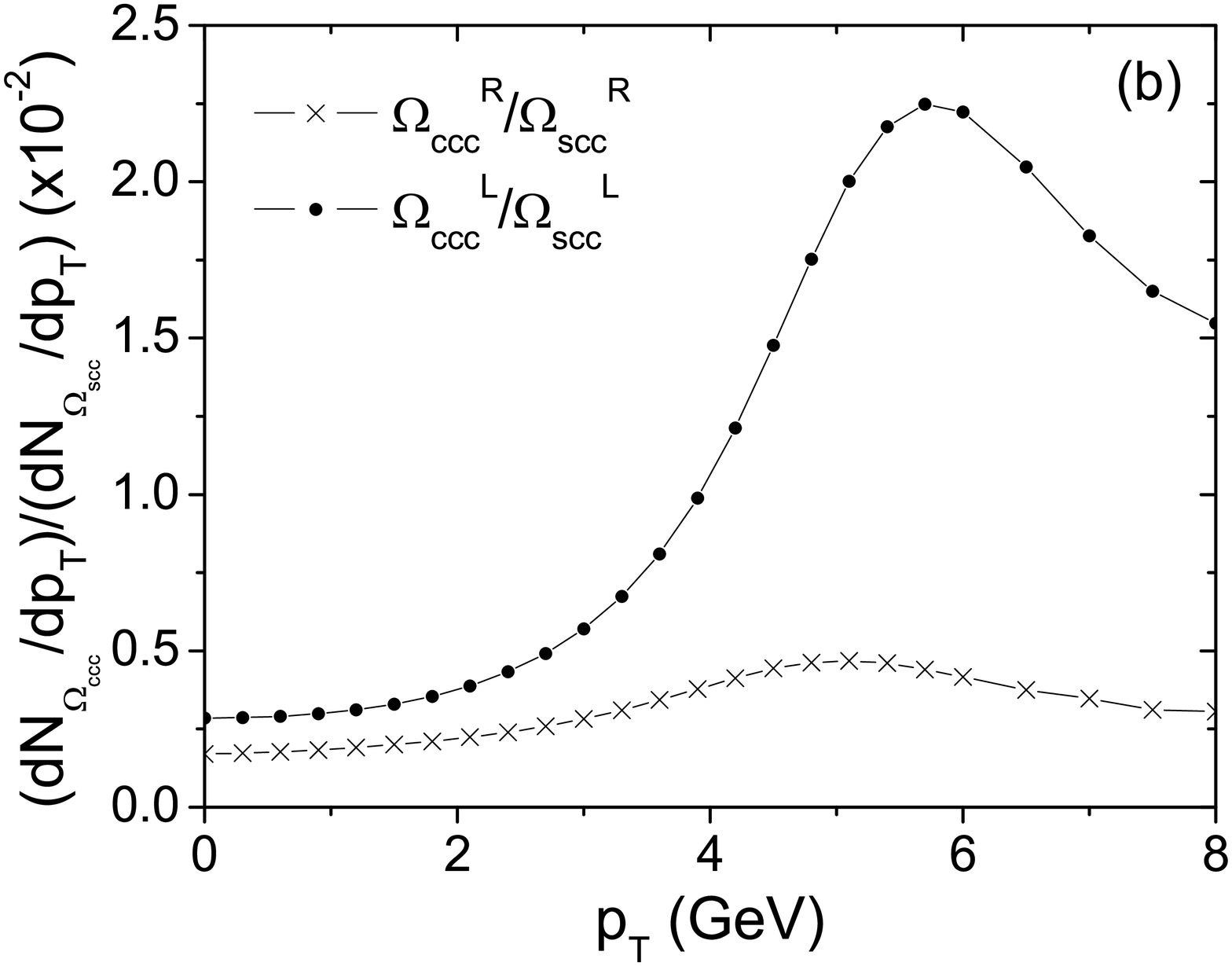}
\includegraphics[width=0.495\textwidth]{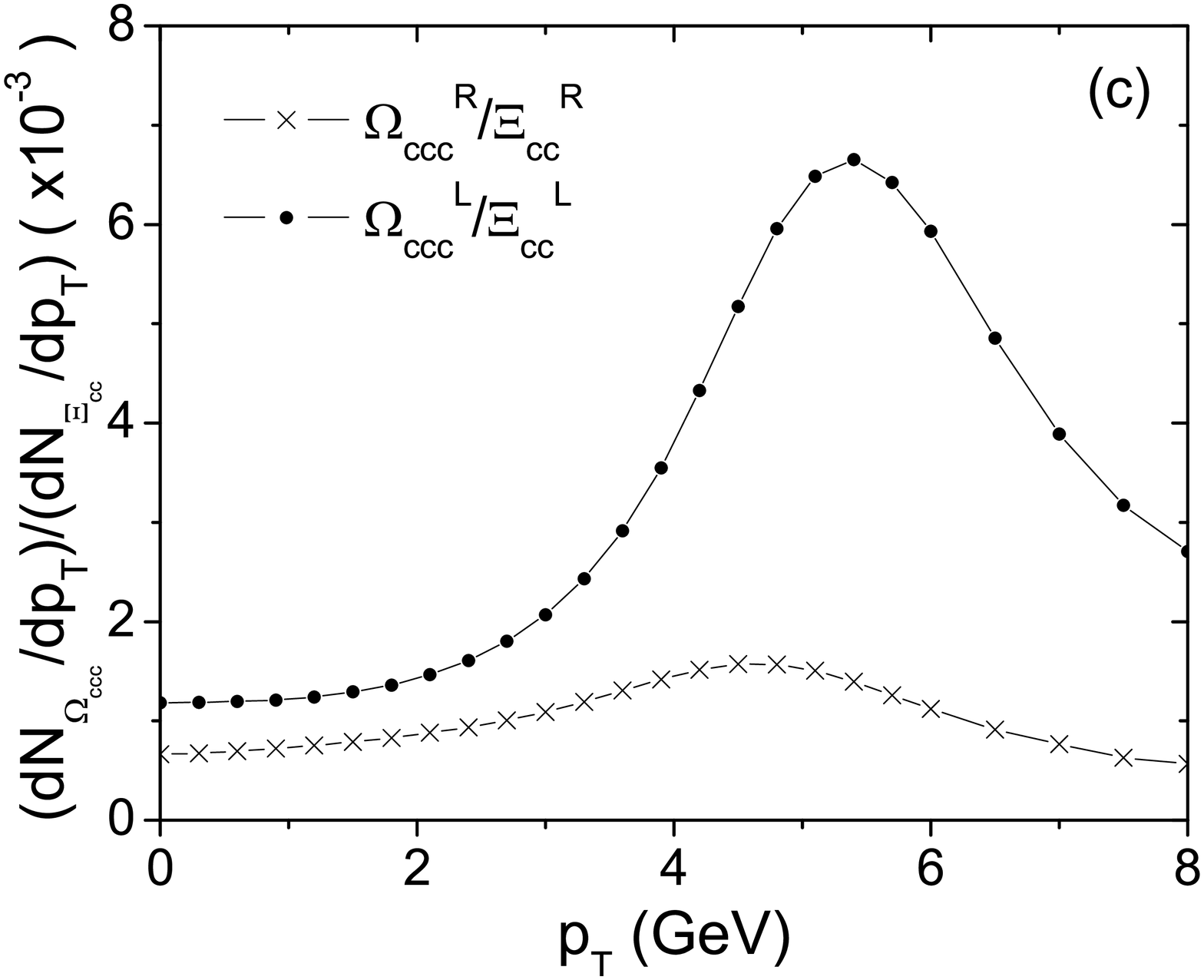}
\includegraphics[width=0.495\textwidth]{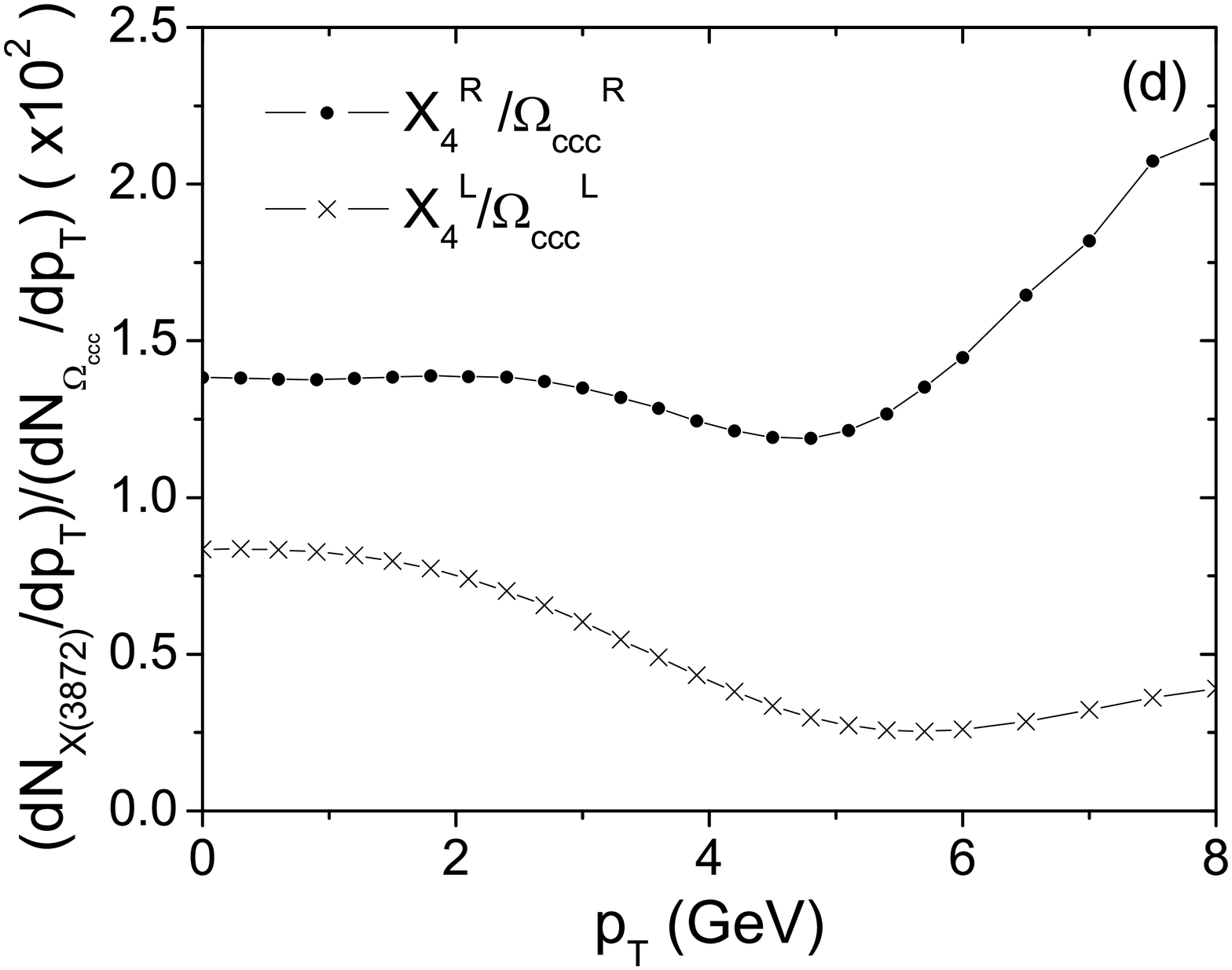}
\includegraphics[width=0.495\textwidth]{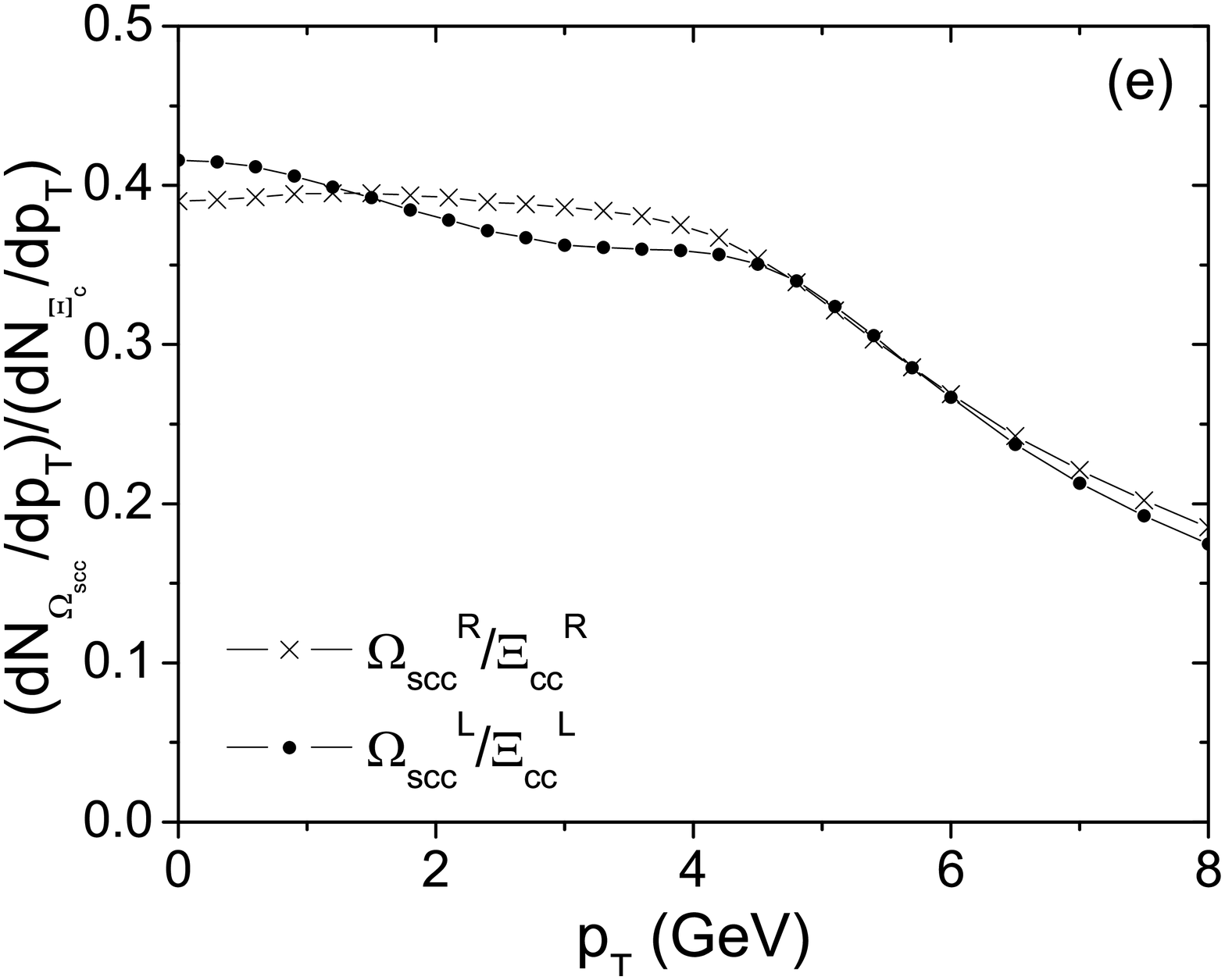}
\includegraphics[width=0.495\textwidth]{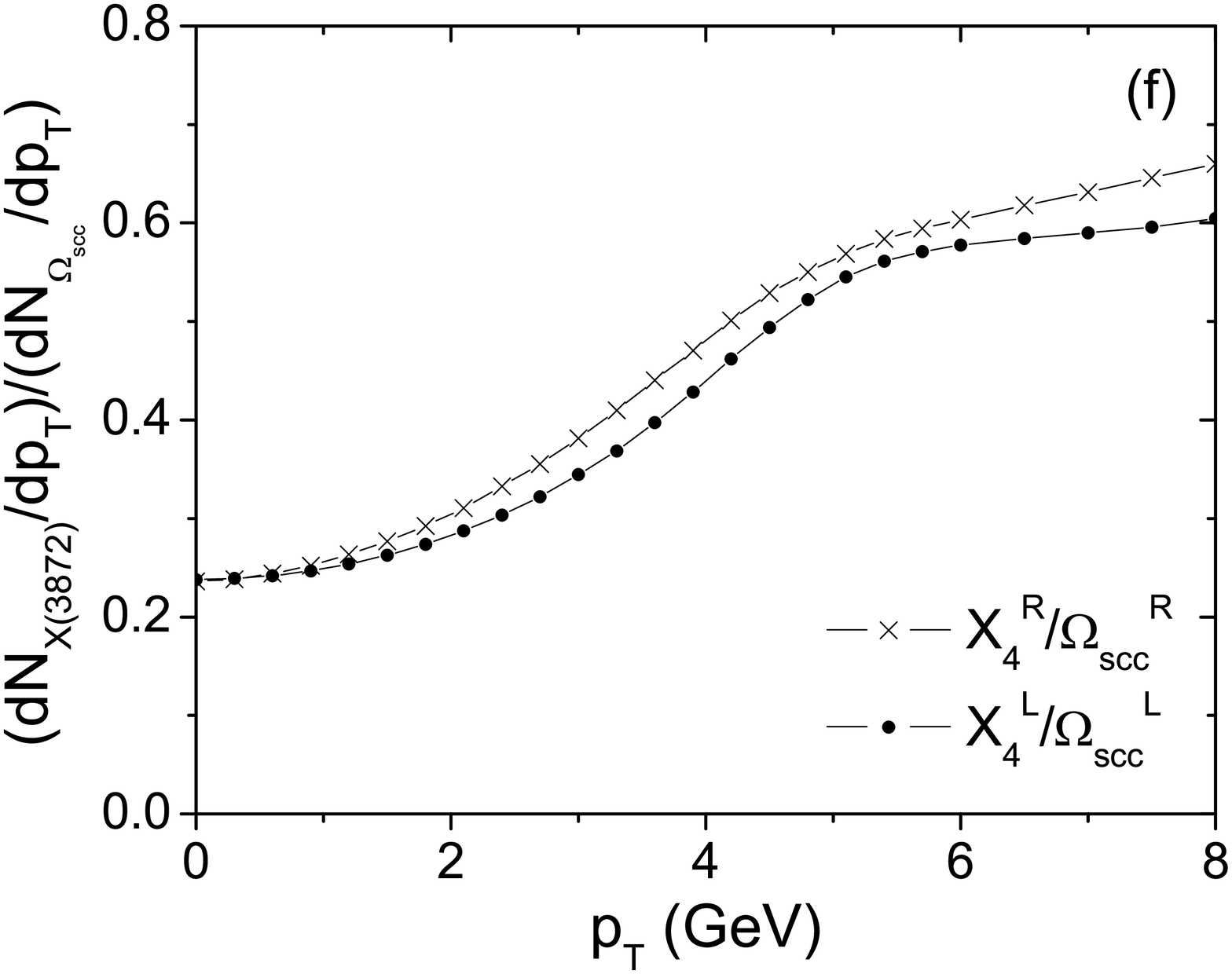}
\end{center}
\caption{Transverse momentum distribution ratios (a) between the
$X(3872)$ and the $\Xi_{cc}$, (b) between the $\Omega_{ccc}$ and
the $\Omega_{scc}$, (c) between the $\Omega_{ccc}$ and the
$\Xi_{cc}$, (d) between the $\Omega_{ccc}$ and the $X(3872)$, (e)
between the $\Xi_{cc}$ and the $\Omega_{scc}$, and (f) between the
$X(3872)$ and the $\Omega_{scc}$ for both RHIC $\sqrt{s_{NN}}$=200
GeV and LHC $\sqrt{s_{NN}}$=2.76 TeV.} \label{MulticharmRatios}
\end{figure}

\end{widetext}

We show in Fig. \ref{MulticharmRatios} transverse momentum
distribution ratios between the $X_4$ and the $\Xi_{cc}$,
$c\bar{c}q\bar{q}/ccq$, between the $\Omega_{ccc}$ and the
$\Omega_{scc}$, $ccc/ccs$, between the $\Omega_{ccc}$ and the
$\Xi_{cc}$, $ccc/ccq$, between the $X_4$ and the $\Omega_{ccc}$,
$c\bar{c}q\bar{q}/ccc$, between the $\Omega_{scc}$ and the
$\Xi_{cc}$, $ccs/ccq$, and between the $X_4$ and the
$\Omega_{scc}$, $c\bar{c}q\bar{q}/ccs$ at both RHIC
$\sqrt{s_{NN}}$=200 GeV and LHC $\sqrt{s_{NN}}$=2.76 TeV. In the
ratio the $\Xi_{cc}$ implies either a $\Xi_{cc}^+$ or a
$\Xi_{cc}^{++}$.

We note from Fig. \ref{MulticharmRatios} that some ratios have a
peak in the intermediate transverse momentum region whereas some
do not. The peak shown in transverse momentum distribution ratios
looks similar to that in the transverse momentum distribution
ratio between an anti-proton and a pion, which is attributable to
a competition between two different hadron production mechanisms
from two different quark distributions; one is an exponential at
low transverse momenta and the other a power law mostly at high
transverse momenta. We find that the peak appearing here in the
transverse momentum distribution ratios between multi-charmed
hadrons is related to the number and type of quark constituents
participating in hadron production.

As introduced in Eq. (\ref{dNldpT}), light quarks in quark-gluon
plasma are assumed to be in thermal equilibrium with an
exponential transverse momentum distribution. On the other hand
charm quarks are not in thermal equilibrium in a system, and
therefore transverse momentum distributions of charm quarks as
shown in Eq. (\ref{dNcdpT}) contain a power law type in addition
to an exponential transverse momentum distribution. For this
reason contributions from charm quarks at higher transverse
momenta prevail that from light quarks; the peak in some ratios
appears when a power law type of charm quark distribution is
involved with a purely exponential type light or strange quarks
whose transverse momentum distributions.

It should be noted that although light quarks also have a power
law type transverse momentum distribution at high transverse
momentum, for such effects to be visible in hadron ratios, the
momentum of a charm quark or a charmed hadron itself should have
an even larger transverse momentum because the transverse momentum
of heavy quark hadrons is mostly dominated by the momentum of the
heavy quarks. Therefore, it is reasonable to consider only light
quarks with an exponential transverse momentum distribution when
we are interested in the transverse momentum distribution of a
heavy quark hadron in the intermediate transverse momentum region.

The transverse momentum distribution ratio between the
$\Omega_{ccc}$ and the $\Omega_{scc}$, $ccc/ccs$ in Fig.
\ref{MulticharmRatios} (b), and that between the $\Omega_{ccc}$
and the $\Xi_{cc}$, $ccc/ccq$ in Fig. \ref{MulticharmRatios} (c)
both showing the peak in the intermediate transverse momentum
region represents the ratio between a charm and a strange or a
light quark in addition to two other common charm quarks. We see
that those two ratios are similar in magnitude, $\sim 10^{-2}$
reflecting that the number of charm quarks in a system is smaller
than that of light quarks by a factor of a few hundreds. The
additional difference in magnitude by a factor 3 between two
ratios is originated from both the number difference between
strange and light quarks in the system, and the heavier mass of
strange quarks compared to that of light quarks.

We also notice that the position of the peak is located at higher
transverse momentum when more heavier quarks are involved; the
peak in Fig. \ref{MulticharmRatios} (b), $c/s$ is shifted to the
higher transverse momentum compared to that in Fig.
\ref{MulticharmRatios} (c), $c/q$. The peak located at the higher
transverse momentum for hadrons with heavier quarks supports the
argument that the momentum of heavy quark hadrons is mostly
carried by heavy quarks due to their heavier mass \cite{Oh:2009zj,
Plumari:2017ntm}.

In addition we further argue that the location of the peak is
closely related to the transverse momentum distribution of the
spectator quark in the ratio. For example, the transverse momentum
of the $\Omega_{ccc}$ is expected to be carried by three charm
quarks equally on average. In case of the $\Omega_{scc}$ with the
same transverse momentum as the $\Omega_{ccc}$, the charm quark in
the $\Omega_{scc}$ will have the larger transverse momentum than
that in the $\Omega_{ccc}$ on average, and therefore makes the
smaller contribution to the transverse momentum distribution of
the $\Omega_{scc}$ from the higher charm quark transverse
momentum, Eq. (\ref{dNcdpT}), or Fig. \ref{dN2dpcT2} compared to
that in the $\Omega_{ccc}$.

Similarly, charm quarks in the $\Xi_{cc}$ with the same transverse
momentum as the $\Omega_{scc}$ has a larger transverse momentum
than those in the $\Omega_{scc}$. Therefore, the light quark in
the $\Xi_{cc}$ has a smaller transverse momentum than the strange
quark in the $\Omega_{scc}$ on average, and thereby the larger
contribution from the light quark transverse momentum distribution
at low transverse momenta, Eq. (\ref{dNldpT}), to the transverse
momentum distribution of the $\Xi_{cc}$ is possible compared to
the strange quark in the $\Omega_{scc}$. On the other hand, at
higher transverse momenta contribution from charm quarks becomes
dominant due to the power law in transverse momentum in their
transverse momentum distributions, Eq. (\ref{dN2dpcT2}). Therefore
there exists both the contribution from the light and charm quark
transverse momentum distribution at intermediate transverse
momentum region, resulting in the peak in the transverse momentum
distribution ratio between the $\Omega_{ccc}$ and the $\Xi_{cc}$.

Moreover, the charm quark in the $\Omega_{scc}$ with the
relatively smaller transverse momentum compared to that in the
$\Xi_{cc}$ together with the strange quark in the $\Omega_{scc}$
with the relatively larger transverse momentum than the light
quark in the $\Xi_{cc}$ has a larger contributions also to the
ratio between the $\Omega_{ccc}$ and the $\Omega_{scc}$ at higher
transverse momenta compared to the case of the $\Xi_{cc}$, and
make the position of the peak a little bit shifted to the higher
transverse momentum compared to the peak in the ratio between the
$\Omega_{ccc}$ and the $\Xi_{cc}$.

We also see that no peaks exist in transverse momentum
distribution ratios between the $\Omega_{scc}$ and the $\Xi_{cc}$,
Fig. \ref{MulticharmRatios} (e) and between the $X(3872)$ and the
$\Omega_{scc}$, Fig. \ref{MulticharmRatios} (f). The ratio in Fig.
\ref{MulticharmRatios} (e) is about $s/q$, and that in
\ref{MulticharmRatios} (f) is about $q\bar{q}/s$ except two
spectator charm quarks. Due to the slight mass difference between
light and strange quarks, 200 MeV, charm quarks in the $\Xi_{cc}$
are expected to have the transverse momentum slightly larger than
that of charm quarks in the $\Omega_{scc}$, resulting in the
smaller transverse momentum for the light quark in the $\Xi_{cc}$
compared to that of the strange quark in the $\Omega_{scc}$ for
all the given transverse momenta. Therefore the contribution from
the light quark in the $\Xi_{cc}$ is always larger than that from
the strange quark in the $\Omega_{scc}$, leading to the decreasing
ratio in Fig. \ref{MulticharmRatios} (e).

The same phenomena takes place in the ratio between the $X(3872)$
and the $\Omega_{scc}$. The transverse momentum of charm quarks in
the $X(3872)$ is smaller than that of charm quarks in the
$\Omega_{scc}$ due to the mass difference between two light quarks
and the strange quark, 100 MeV, making each light quark to have
the smaller transverse momentum than the strange quark in the
$\Omega_{scc}$. Therefore, the contribution from two light quarks
in the $X(3872)$ is always larger than that from the strange quark
in the $\Omega_{scc}$ for all the given transverse momenta,
leading to the increasing ratio in Fig. \ref{MulticharmRatios}
(f).

It should be also noted that in Fig. \ref{MulticharmRatios} (e)
the ratio is expected to stop decreasing and begin to increase
while the ratio in Fig. \ref{MulticharmRatios} (f) is expected to
stop increasing and begin to decrease at very high transverse
momentum region eventually. As has been mentioned above, light
quarks are assumed in thermal equilibrium with only an exponential
transverse momentum distribution. When light quarks actually have
a power law transverse momentum distribution, the ratio between
heavy hadrons containing light quarks where the remaining factors
involve only light quarks like that in Fig. \ref{MulticharmRatios}
(f), can show peaks but at very high transverse momentum because
the power law effects of light quarks are turned on at only very
high transverse momentum in hadrons with heavy quarks.

The transverse momentum distribution ratio between the $X(3872)$
and the $\Omega_{ccc}$, Fig. \ref{MulticharmRatios} (d) leaving
quarks of $q\bar{q}/c$ in the ratio also shows no peaks. We find
that there can exist the peak when the inverse of the ratio, the
transverse momentum distribution ratio between the $\Omega_{ccc}$
and the $X(3872)$, $c/\bar{q}/q$ is taken. The light quark in the
$X(3872)$ having the much smaller transverse momentum than charm
quarks in the $\Omega_{ccc}$ has a large contribution to the
transverse momentum distribution of the $X(3872)$ at low
transverse momenta, resulting in the decreasing ratio in Fig.
\ref{MulticharmRatios} (d). However, at higher transverse momenta
the contribution from charm quarks becomes dominant, resulting in
the increasing ratio in Fig. \ref{MulticharmRatios} (d) with
increasing transverse momenta.

In that sense the transverse momentum distribution ratio between
the $X(3872)$ and the $\Xi_{cc}$, given in Fig.
\ref{MulticharmRatios} (a), is noticeable since it has a peak in
the intermediate transverse momentum region even though the ratio
is again about light quarks, $q\bar{q}/q$. The transverse momentum
distribution ratio between the $X(3872)$ and the $\Xi_{cc}$ is
similar to the ratio between anti-protons and pions, or that
between the $\Lambda_c$ and the $D^0$ since those ratios leave the
same number of quarks in the ratio, $q\bar{q}/q$. The reason for
the peak appearing in Fig. \ref{MulticharmRatios} (a) is that
quarks in the ratio are quarks of the same type contrary to those
in Fig. \ref{MulticharmRatios} (f). As shown in Table
\ref{yieldratios} yield ratios between the $X(3872)$ and the
$\Xi_{cc}$ are at most 0.11 at RHIC and 0.12 at LHC in the
statistical hadronization model when the isospin is taken into
account. We see in Fig. \ref{MulticharmRatios} (a) that the ratio
increases up to about 0.20 at both RHIC and LHC, no significant
enhanced production of the heavier $X(3872)$ meson compared to the
$\Xi_{cc}$ baryon in the intermediate transverse momenta.

\begin{widetext}

\begin{figure}[!h]
\begin{center}
\includegraphics[width=0.495\textwidth]{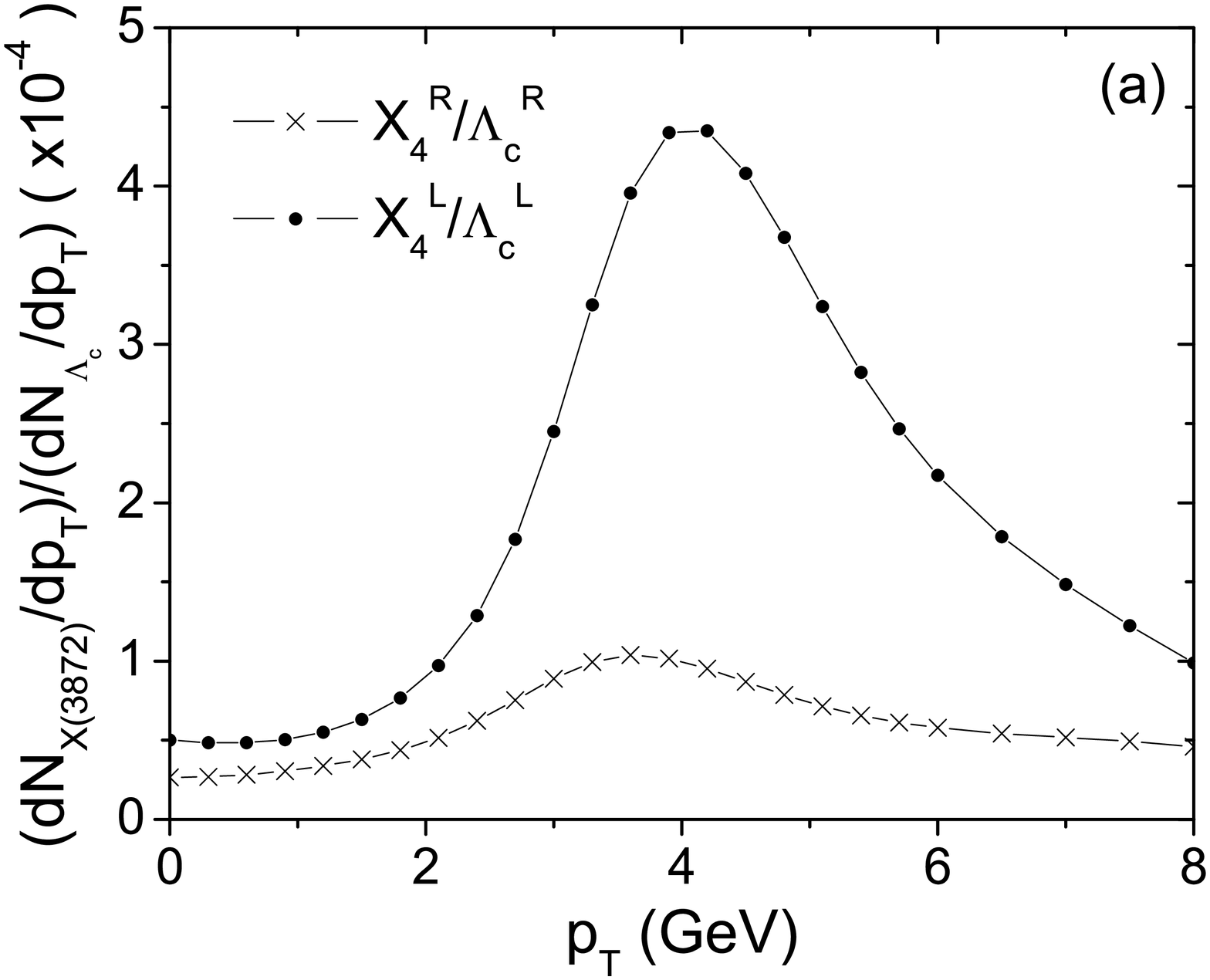}
\includegraphics[width=0.495\textwidth]{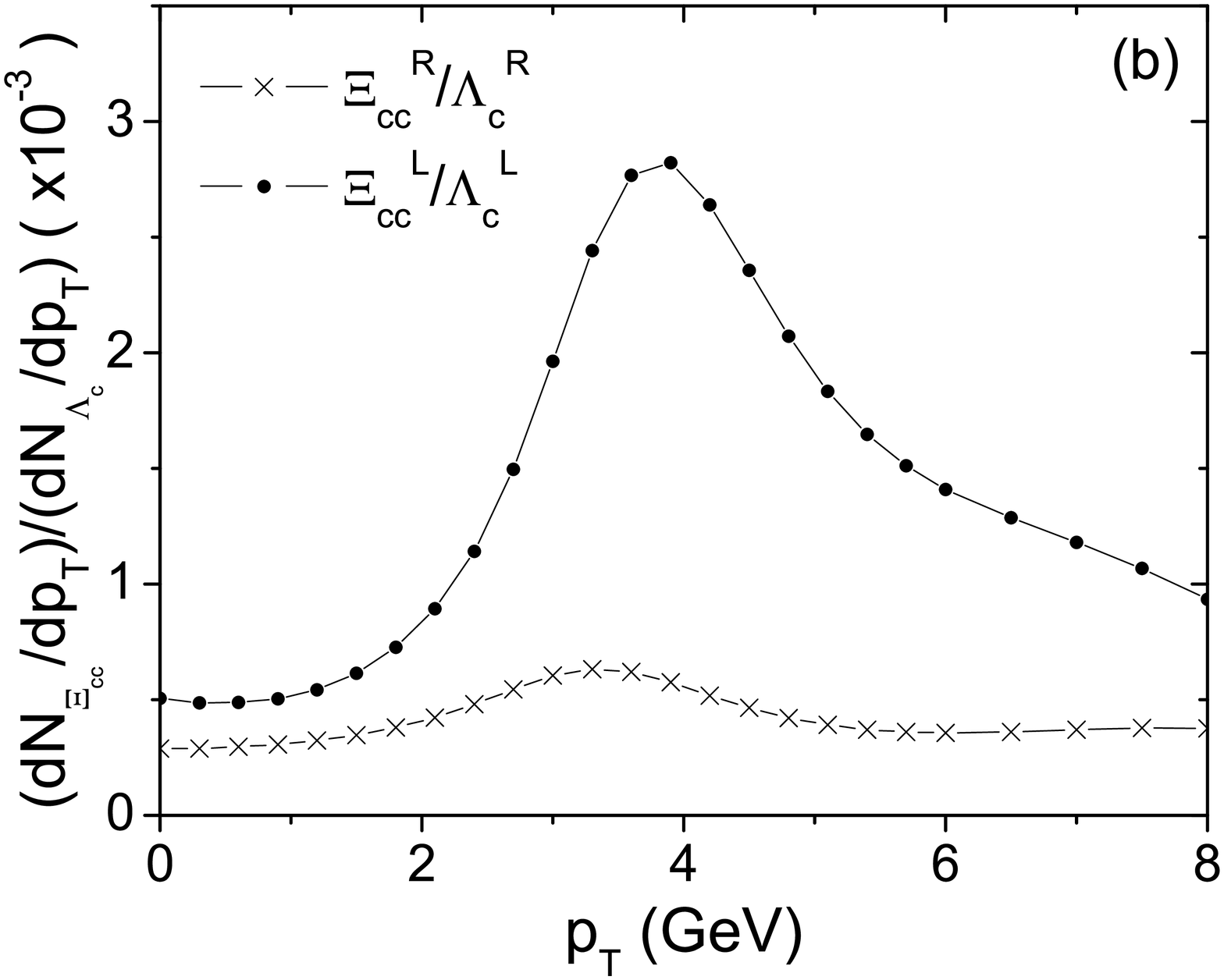}
\includegraphics[width=0.495\textwidth]{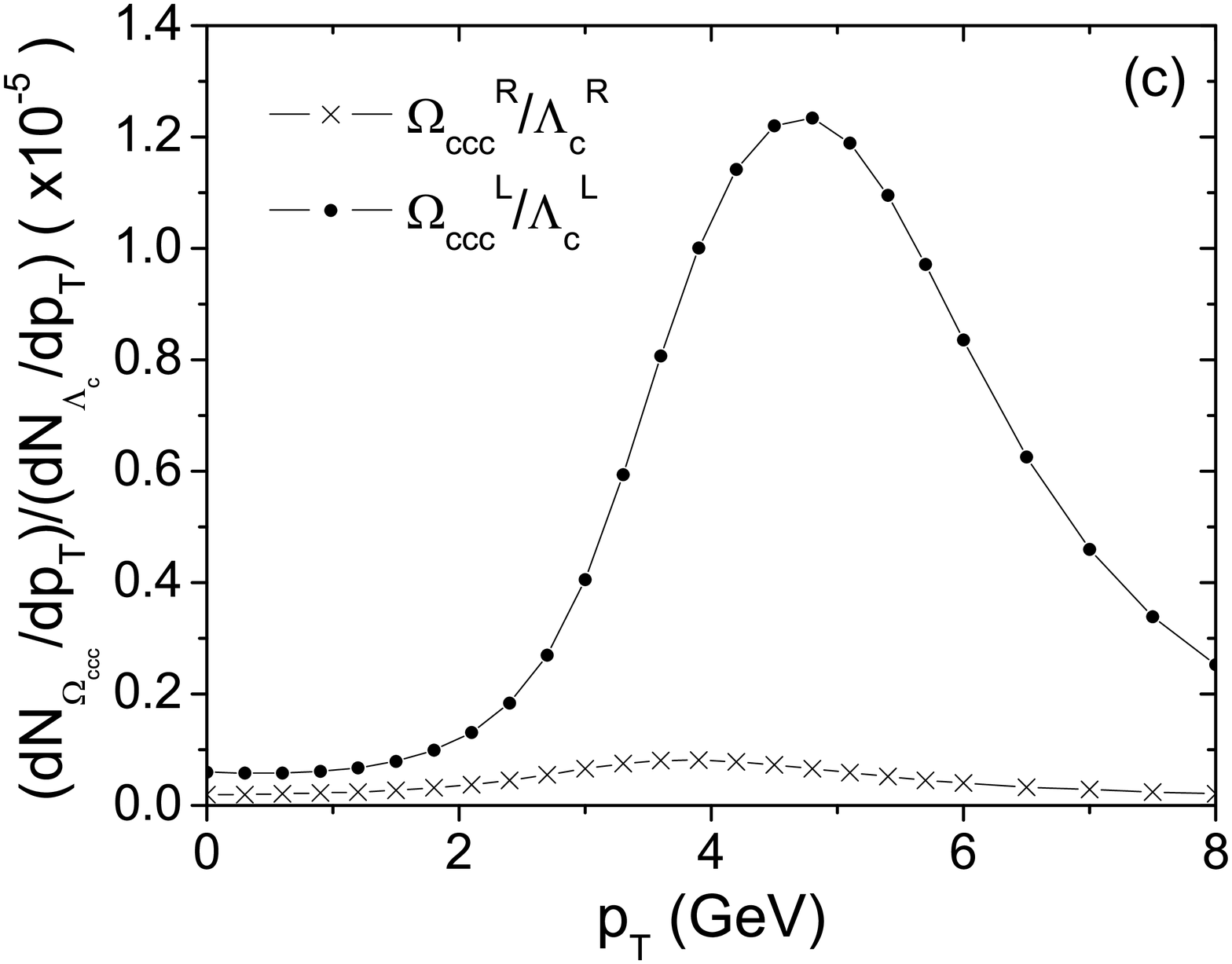}
\includegraphics[width=0.495\textwidth]{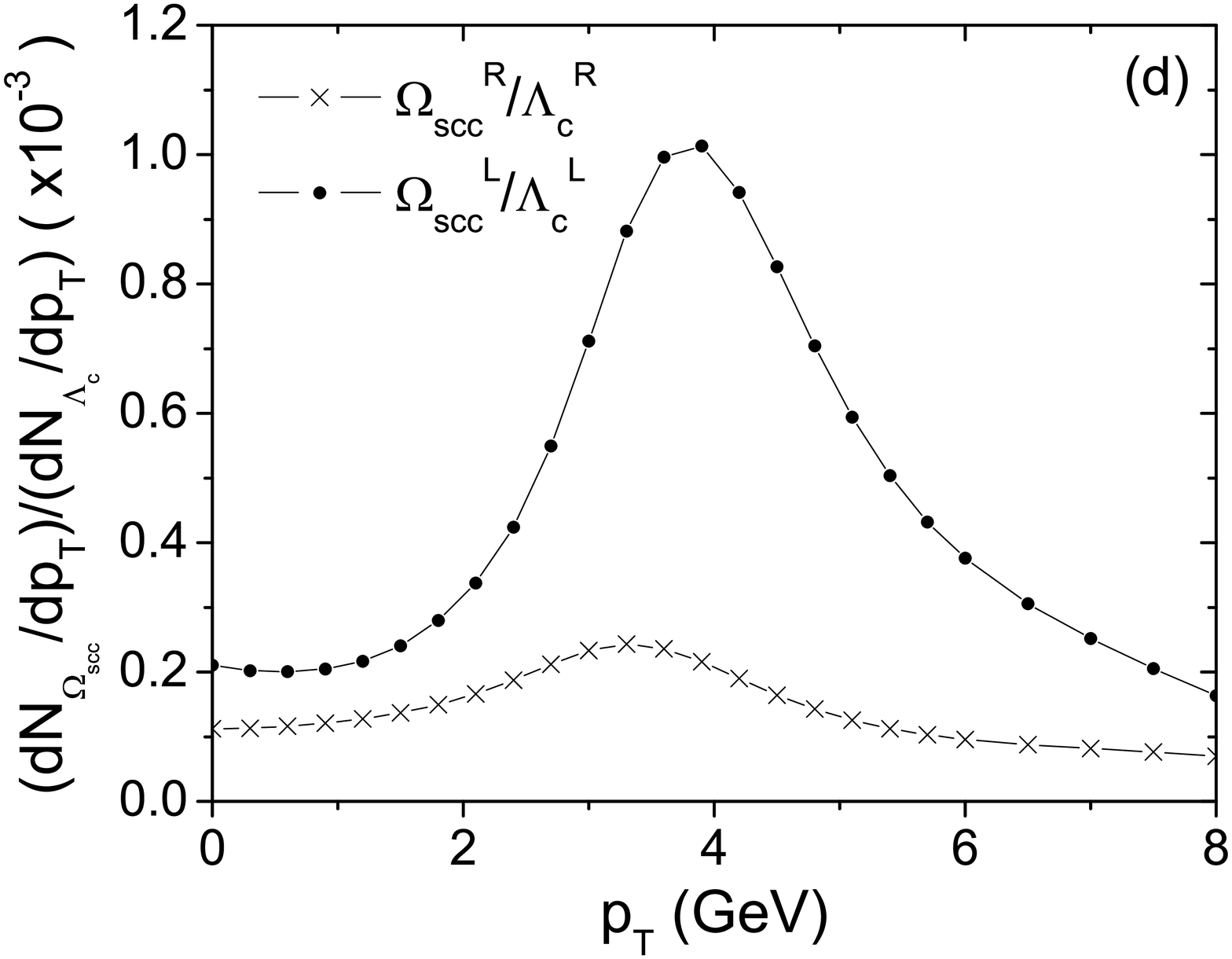}
\end{center}
\caption{Transverse momentum distribution ratios (a) between the
$X(3872)$ and the $\Lambda_c$, (b) between the $\Xi_{cc}$ and the
$\Lambda_c$, (c) between the $\Omega_{ccc}$ and the $\Lambda_c$,
and (d) between the $\Omega_{scc}$ and the $\Lambda_c$ at both
RHIC $\sqrt{s_{NN}}$=200 GeV and LHC $\sqrt{s_{NN}}$=2.76 TeV.}
\label{singlecharmRatios}
\end{figure}

\end{widetext}

Moreover, as we see in Fig. \ref{MulticharmRatios} (a), (e) and
(f), transverse momentum distribution ratios between the $X(3872)$
and the $\Xi_{cc}$, between the $\Omega_{scc}$ and the $\Xi_{cc}$,
and between the $X(3872)$ and the $\Omega_{scc}$ at RHIC are very
similar to those at LHC. Since the same function for the light
quark transverse momentum distribution, Eq. (\ref{dNldpT}) with
the same effective temperature at both RHIC and LHC has been
introduced in the analysis, ratios at RHIC and LHC obtained after
cancelling two spectator charm quarks should be somehow similar to
each other. In other words, the difference of the transverse
momentum distribution ratios at RHIC and LHC shown in Figs.
\ref{MulticharmRatios} (b), (c) and (d) must be originated from
explicitly different transverse momentum distributions of charm
quarks at RHIC and LHC, Eq. (\ref{dNcdpT}).

\subsection{Transverse momentum distribution ratios between
a multi-charmed hadron and a $\Lambda_c$}

We compare the transverse momentum distribution of multi-charmed
hadrons with that of the singly-charmed hadron, the $\Lambda_c$.
For the $\Lambda_c$ transverse momentum distribution we have
included the contribution of the $\Lambda_c$ production by
fragmentation as well as feed-down contributions from
$\Sigma_c(2455)$, $\Sigma_c(2520)$, $\Lambda_c(2595)$, and
$\Lambda_c(2625)$ baryons. We show in Fig. \ref{singlecharmRatios}
four transverse momentum distribution ratios, (a) between the
$X(3872)$ and the $\Lambda_c$, (b) between the $\Xi_{cc}$ and the
$\Lambda_c$, (c) between the $\Omega_{ccc}$ and the $\Lambda_c$,
and (d) between the $\Omega_{scc}$ and the $\Lambda_c$ for both
RHIC, $\sqrt{s_{NN}}$=200 GeV and LHC, $\sqrt{s_{NN}}$=2.76 TeV.

\begin{widetext}

\begin{figure}[!t]
\begin{center}
\includegraphics[width=0.495\textwidth]{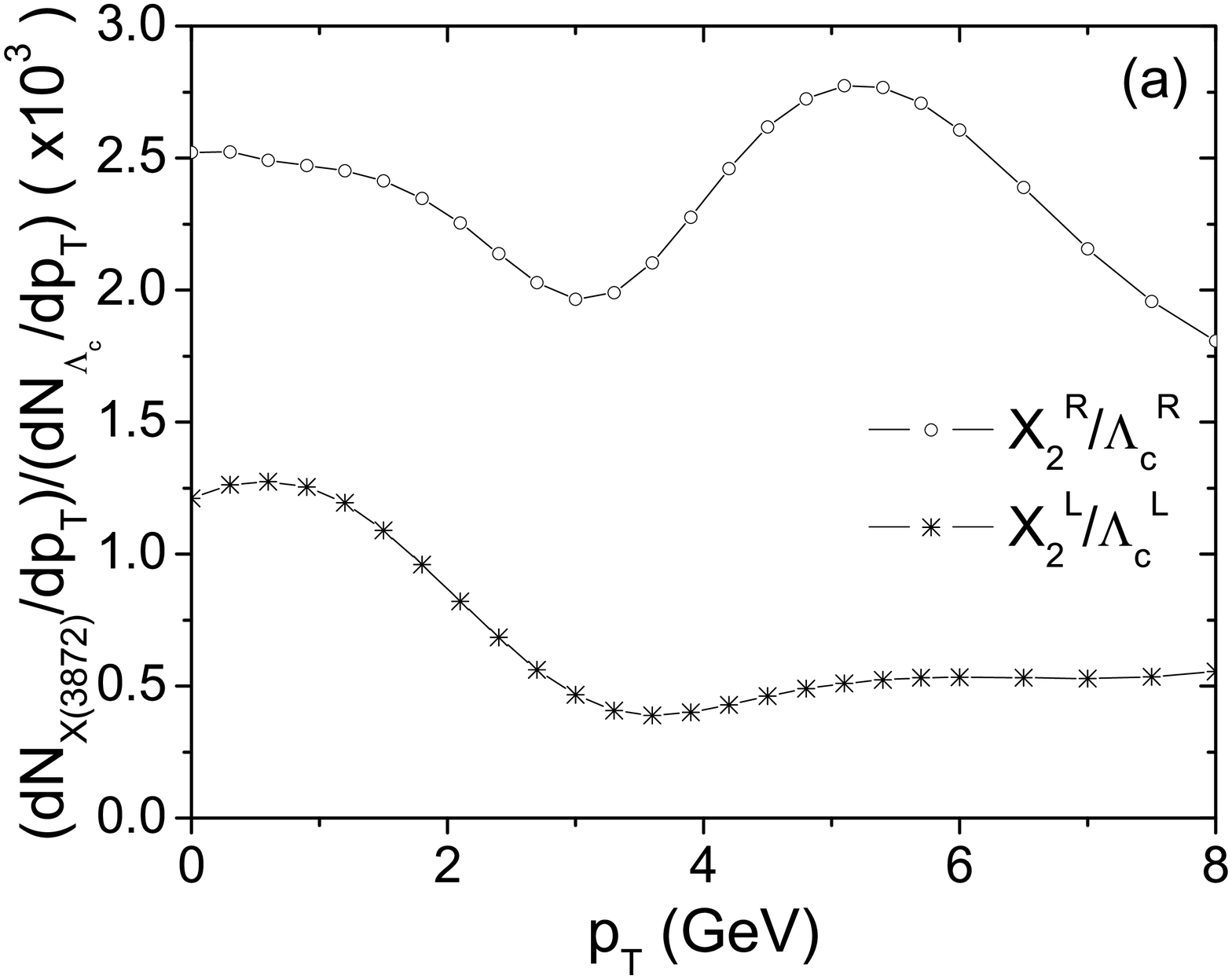}
\includegraphics[width=0.495\textwidth]{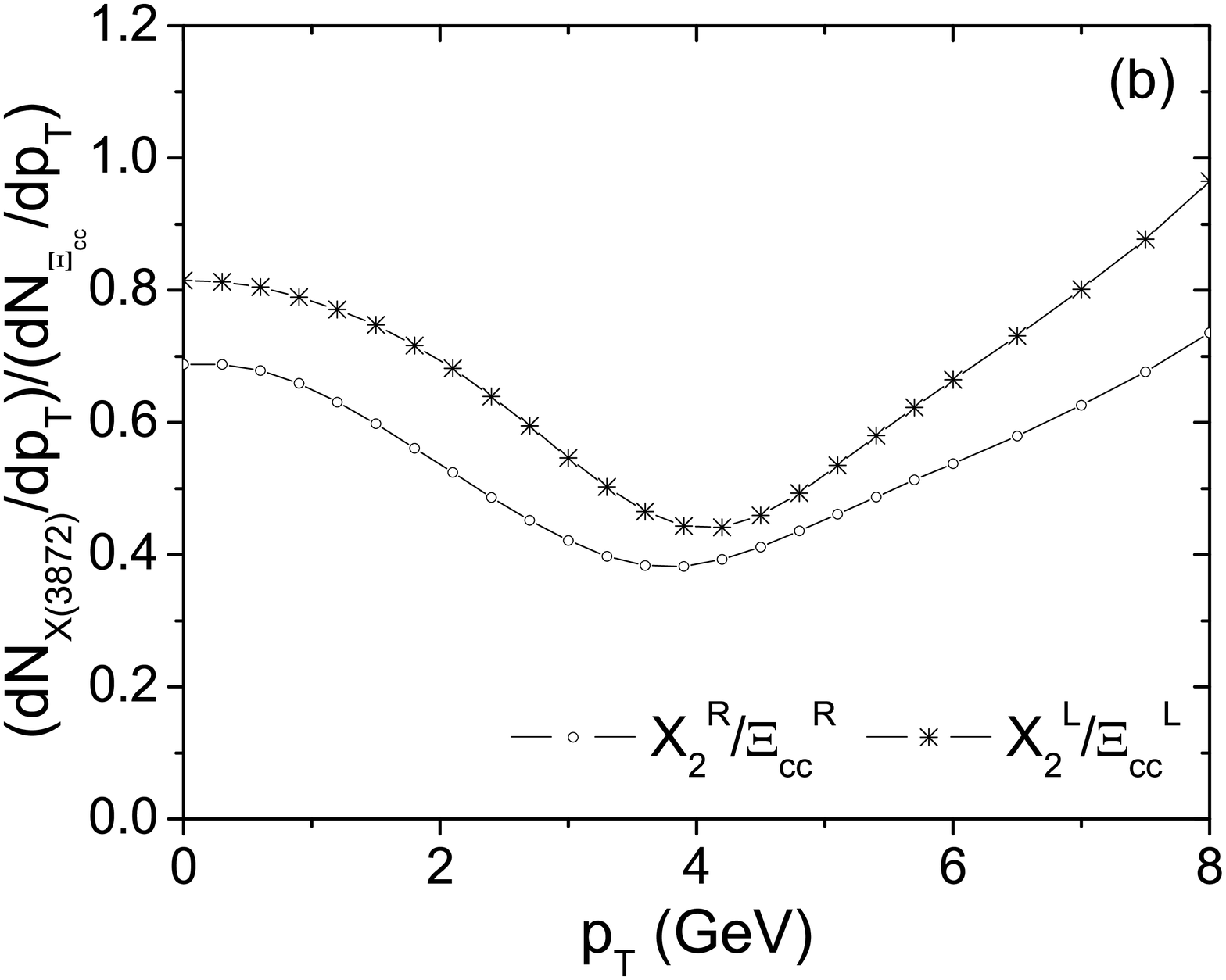}
\includegraphics[width=0.495\textwidth]{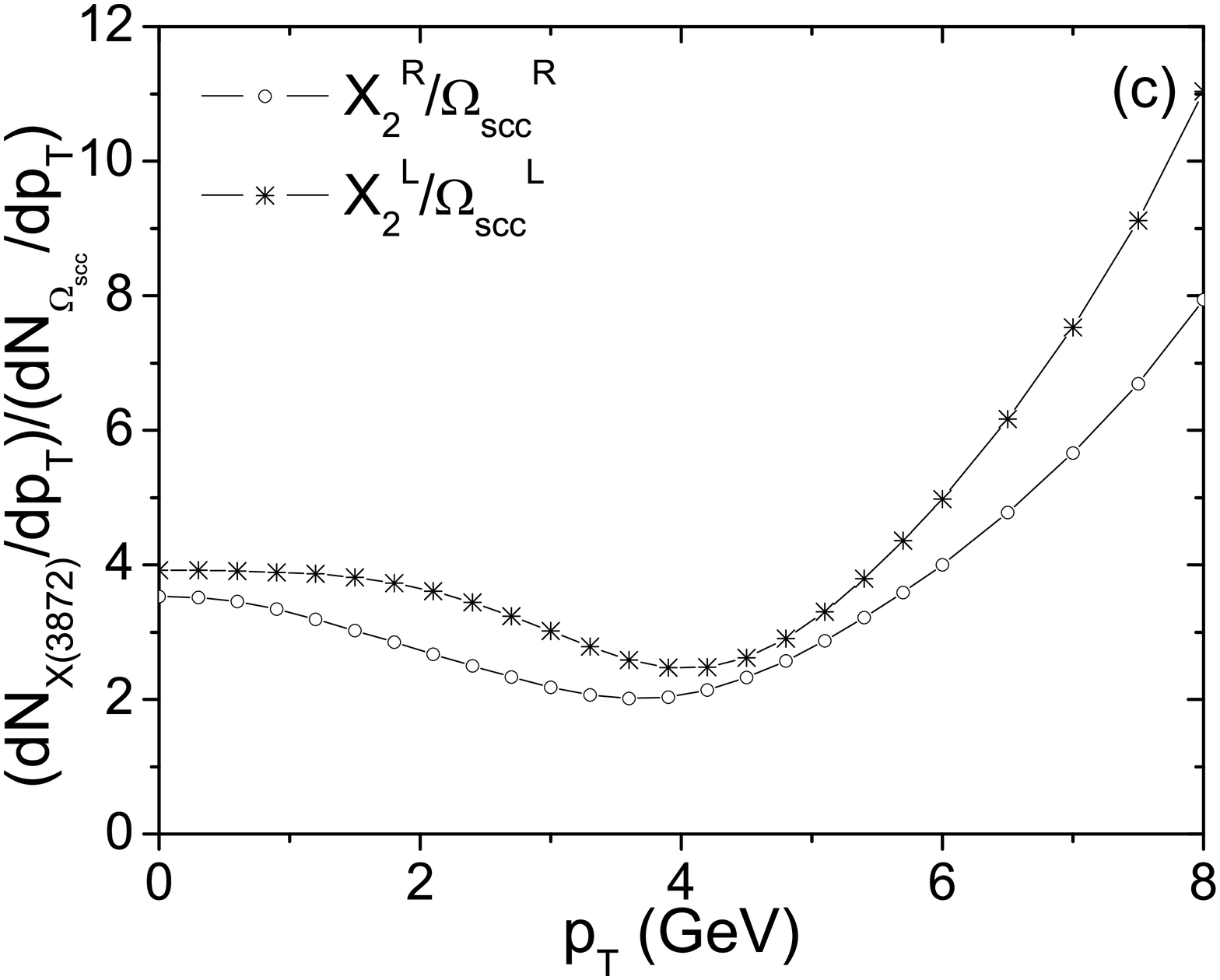}
\includegraphics[width=0.495\textwidth]{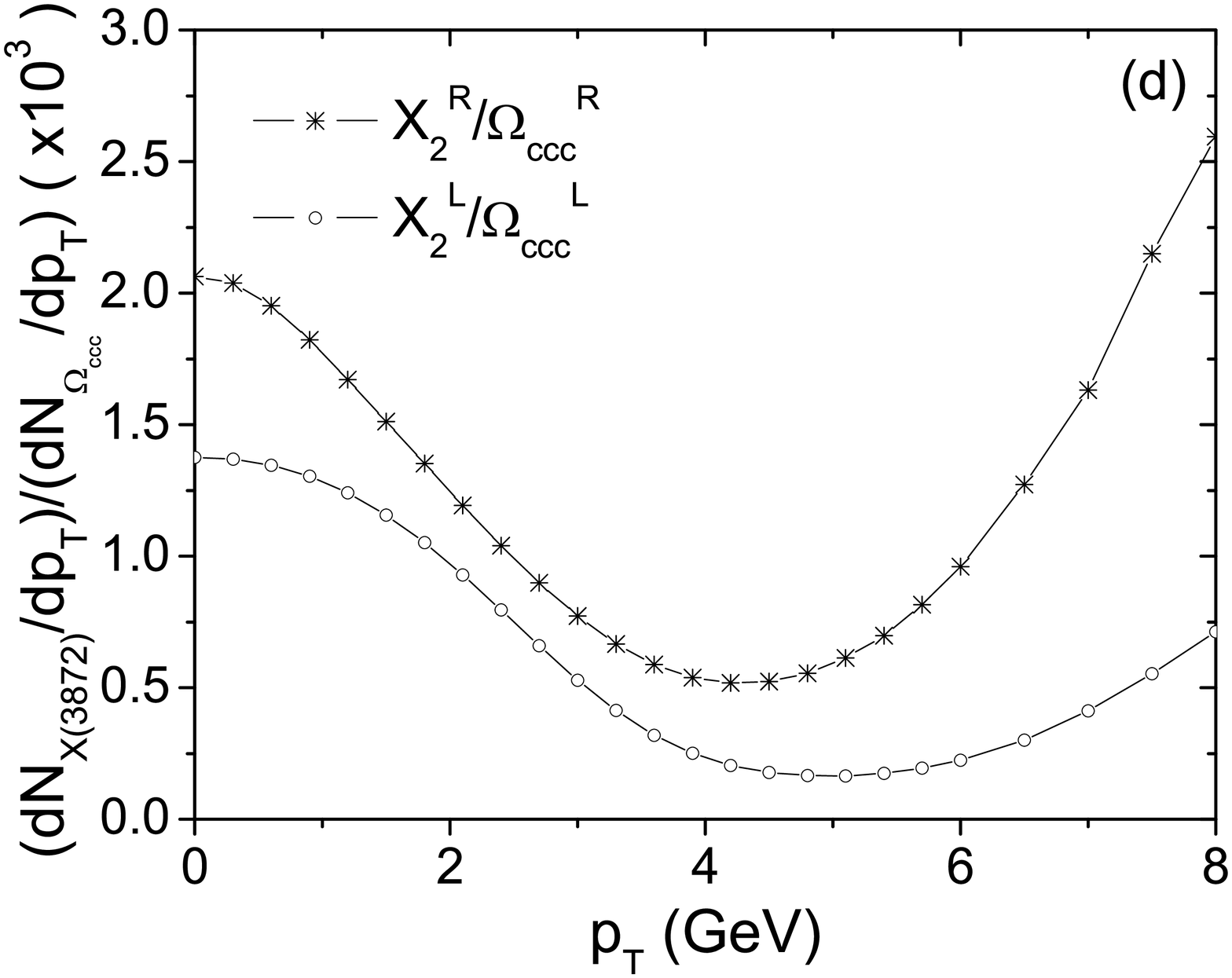}
\end{center}
\caption{Transverse momentum distribution ratios (a) between the
$X(3872)$ in a two-quark state, the $X_2$ and the $\Lambda_c$, (b)
between the $X_2$ and the $\Xi_{cc}$, (c) between the $X_2$ and
the $\Omega_{scc}$, and (d) between the $X_2$ and the
$\Omega_{scc}$ for both RHIC $\sqrt{s_{NN}}$=200 GeV and LHC
$\sqrt{s_{NN}}$=2.76 TeV.} \label{X2Ratios}
\end{figure}

\end{widetext}

As we see in Fig. \ref{singlecharmRatios}, the ratio is much
smaller than unity, again reflecting the small possibility to
coalesce more charm quarks to form a multi-charmed hadron. Since
the number of charm quarks are smaller than that of light quarks
by an order of two, the ratio between the $X(3872)$ and the
$\Lambda_c$ is also smaller than that between the $X(3872)$ and
the $\Xi_{cc}$ by the same order. Nevertheless we still see peaks
appearing in the intermediate transverse momentum region, but at
lower transverse momentum about 4 GeV.

We have argued that the peak can appear for the ratio involving
both light quarks in thermal equilibrium with an exponential
transverse momentum distribution and charm quarks with a power law
type transverse momentum distribution in addition to an
exponential transverse momentum distribution. We have also found
that the peak appears in the ratio involving pure light quarks
with the same kind but different numbers in the numerator and the
denominator, e.g., $q\bar{q}/q$.

We further argue that a peak appears in the ratio involving charm
and light quarks, especially when a remaining charm quark is in
the numerator and a light quark remains in the denominator, e.g.,
$c/q$. As shown in Fig. \ref{MulticharmRatios} the peak appears in
ratios, $c/s$ (b) and $c/q$ (c) except the peak in the ratio,
$q\bar{q}/q$ (a). No peak appears for the ratio, $q/c$ (d), $s/q$
(e), and $q\bar{q}/s$ (f). We see that all the ratios shown in
Fig. \ref{singlecharmRatios} involve at least one charm quark in
the numerator and one light quark in the denominator,
$\bar{q}\bar{c}/q$ (a), $c/q$ (b), $cc/qq$ (c), and $cs/qq$ (d),
and therefore we find that the peak always appears in the ratio
between a multi-charm hadron and a $\Lambda_c$.

Transverse momentum distribution ratios shown in Fig.
\ref{singlecharmRatios} (a), (b) and (d) look very similar in both
shape and magnitude $\sim10^{-3}$; three ratios represents
$c\bar{c}q\bar{q}/cqq$, $ccq/cqq$, and $ccs/cqq$, respectively. If
we neglect spectator quarks we see ratios $\bar{c}\bar{q}/q$,
$c/q$, and $cs/qq$. The inclusion of one more light quark in the
numerator, $c\bar{c}q\bar{q}$ not only suppresses more the ratio
$c\bar{c}q\bar{q}/cqq$ to $\sim10^{-4}$, and also broadens the
peak in Fig. \ref{singlecharmRatios} (a) compared to other two
ratios. The same exponential transverse momentum distribution and
200 MeV mass difference between light and strange quarks give the
similar ratios, $c/q$ and $cs/qq$ as shown in Fig
\ref{singlecharmRatios} (b) and (d).

As we see in Fig. \ref{singlecharmRatios} (c), the ratio between
the $\Omega_{ccc}$ and the $\Lambda_c$, $ccc/ccq$ is smaller than
other ratios due to one more charm quark in the numerator, the
$\Omega_{ccc}$, and the peak in the $ccc/ccq$ ratio is shifted to
the higher transverse momentum compared to other three ratios. If
we compare all the ratios involving the $\Omega_{ccc}$, Fig.
\ref{MulticharmRatios} (b), (c), and Fig. \ref{singlecharmRatios}
(c), we see that the ratio with the doubly-charmed hadron, the
$\Xi_{cc}$ and the $\Omega_{scc}$ has a peak at the higher
transverse momentum. We also find that the ratio with the
$\Omega_{ccc}$ has the broader peak than other ratios. As has been
pointed out \cite{Oh:2009zj} this is originated from the heavy
mass of charm quarks. Since the momentum of heavy quark hadrons is
mostly carried out by heavy quarks, it is expected that the
transverse momentum distribution would be independent of the
transverse momentum at the limit of the infinite quark mass. By
this reason, the ratio of the $\Omega_{ccc}$ with the
doubly-charmed baryon has the broader peak at the higher
transverse momentum than the ratio with the $\Lambda_c$.

\subsection{Transverse momentum distribution ratios between
an $X_2$ and a charmed hadron}

Using the same transverse momentum distribution of charmed hadrons
shown in Fig. \ref{MultiCharm_pT} we also evaluate transverse
momentum distribution ratios between the $X(3872)$ meson in a
two-quark state and the charmed baryon, the $\Lambda_c$,
$\Xi_{cc}$, $\Omega_{scc}$, and $\Omega_{ccc}$. These are ratios
between the meson and the baryon, or those between two and three
quarks. The ratio between the $X_2$ and the $\Lambda_c$ is the
ratio, $\bar{c}/qq$, that between the $X_2$ and the $\Xi_{cc}$ is
the ratio, $\bar{c}/qc$, that between the $X_2$ and the
$\Omega_{scc}$ is the ratio, $\bar{c}/sc$, and that between the
$X_2$ and the $\Omega_{ccc}$ is the ratio, $\bar{c}/cc$ after
cancelling the spectator quarks. We show these transverse momentum
distribution ratios in Fig. \ref{X2Ratios}.

\begin{figure}[!t]
\begin{center}
\includegraphics[width=0.51\textwidth]{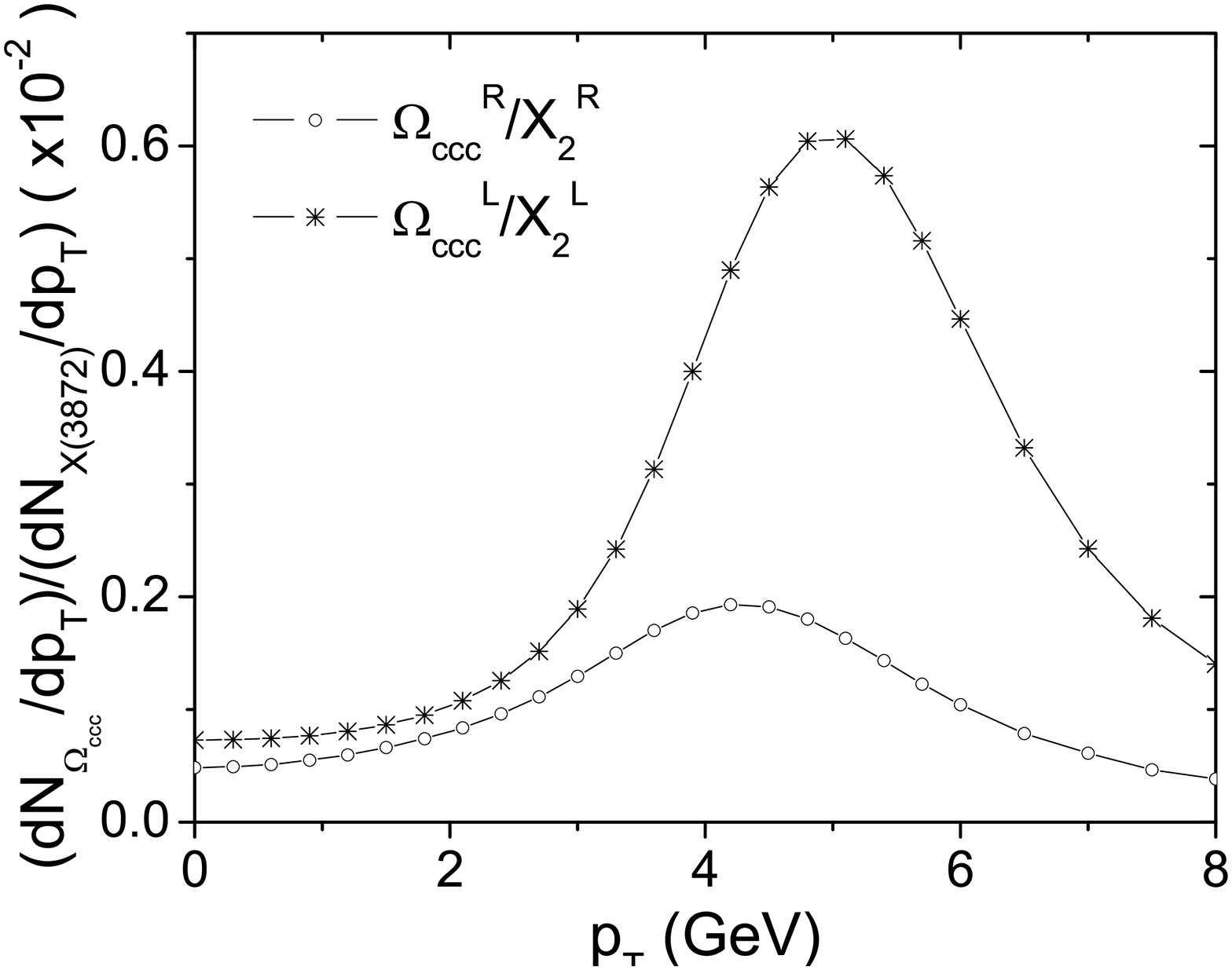}
\end{center}
\caption{The transverse momentum distribution ratio between the
$\Omega_{ccc}$ and the $X(3872)$ in a two-quark state,
$ccc/c\bar{c}$.} \label{OmegaccctoX2}
\end{figure}

As shown in Fig. \ref{X2Ratios}, the ratios between the $X_2$ and
charmed baryons are completely different from those between the
$X(3872)$ in a four-quark state and charmed baryons, Figs
\ref{MulticharmRatios} and \ref{singlecharmRatios}. We do not see
any peak in the ratio between the $X_2$ and the charmed baryon in
Fig. \ref{X2Ratios}. Moreover, the ratio shown in Fig.
\ref{X2Ratios} (b), (c), and (d) looks like the upside down of the
ratio with the peak. We actually confirm that there exist peaks
when we evaluate transverse momentum distribution ratios between
the multi-charmed hadron and the $X_2$; the ratio between the
$\Xi_{cc}$ and the $X_2$, $cq/\bar{c}$, that between the
$\Omega_{scc}$ and the $X_2$, $sc/\bar{c}$, and that between the
$X_2$ and the $\Omega_{ccc}$, $cc/\bar{c}$. Even though we obtain
the similar plot as shown in Fig. \ref{MulticharmRatios} (b) if we
evaluate the transverse momentum distribution ratio between the
$\Omega_{scc}$ and the $X_2$, we show the transverse momentum
distribution ratio between the $X_2$ and the $\Omega_{scc}$ as
shown in Fig. \ref{X2Ratios} (c) in order to make the comparison
easier, i.e., in order to consider only the ratio of the $X(3872)$
to any charmed hadrons. Nevertheless, it is interesting to observe
the peak in the ratio between the $\Omega_{ccc}$ and the $X_2$ as
shown in Fig. \ref{OmegaccctoX2} similar to that in the ratio
between the anti-proton and the pion. It must be due to similar
quark contents in two ratios, $\bar{q}\bar{q}/q$ for the
$\bar{p}/\pi$ and $cc/\bar{c}$ for the $\Omega_{ccc}/X_2$.

\begin{figure}[!t]
\begin{center}
\includegraphics[width=0.49\textwidth]{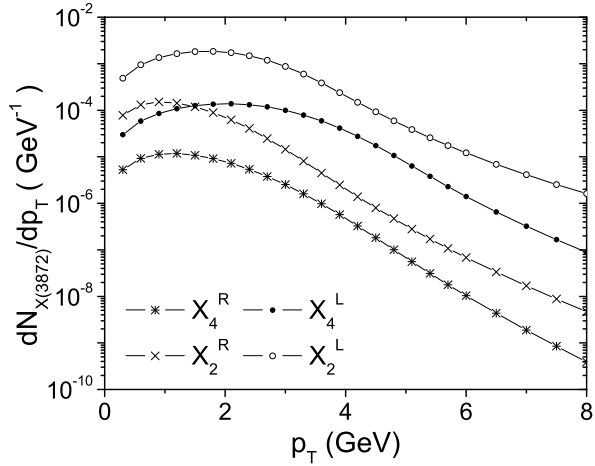}
\end{center}
\caption{Transverse momentum distributions of both the $X(3872)$
in a four-quark sate, $X_4$ and that in a two-quark state, $X_2$
for $\sqrt{s_{NN}}$=200 GeV at RHIC and $\sqrt{s_{NN}}$=2.76 TeV
at LHC.} \label{dNdpTX2X4}
\end{figure}

As expected from the transverse momentum distribution ratio
between the $X(3872)$ and various charmed hadrons, the transverse
momentum distribution of the $X(3872)$ meson in a two-quark state
is quite different from that of the $X(3872)$ meson in a
four-quark state. Therefore, we consider that we can identify
whether the $X(3872)$ meson is composed of four quarks or two
quarks by measuring the transverse momentum distribution ratio
between the $X(3872)$ and various charmed baryons. We also show
transverse momentum distributions, $dN_{X(3872)}/dp_T$ of both the
$X(3872)$ in a four-quark sate and that in a two-quark state in
Fig. \ref{dNdpTX2X4}. Recently, transverse momentum spectra of the
$X(3872)$ cross section for Pb-Pb and Kr-Kr collisions at
$\sqrt{s}=5$ TeV have been predicted in the statistical
hadronization model \cite{Andronic:2019wva}. We hope that we can
compare directly the transverse momentum distribution of the
$X(3872)$ yield obtained here to that measured in relativistic
heavy ion collision experiments as well as that obtained in the
statistical hadronization model in the near future, and that we
can identify the quark structure of the $X(3872)$ meson.

\section{Conclusion}

We have studied the production of multi-charmed hadrons by
recombination in relativistic heavy ion collisions by focusing on
the production of $\Xi_{cc}$, $\Xi_{cc}^*$, $\Omega_{scc}$,
$\Omega_{scc}^*$, $\Omega_{ccc}$ baryons and $X(3872)$ mesons. We
first pay attention to the yield in heavy ion collisions, and have
estimated that of multi-charmed hadrons mentioned above at
chemical freeze-out in both the statistical and coalescence model.
We have also discussed the various yield ratio between
multi-charmed hadrons.

Secondly we focus on the transverse momentum distribution in heavy
ion collisions, and have evaluated that of multi-charmed hadrons,
the $\Xi_{cc}$, $\Xi_{cc}^*$, $\Omega_{scc}$, $\Omega_{scc}^*$,
$\Omega_{ccc}$ baryon and the X(3872) meson at mid-rapidity in the
coalescence model. We have also obtained transverse momentum
distribution ratios between multi-charmed hadrons, especially
transverse momentum distribution ratios related to the $X(3872)$
meson, a meson/baryon ratio similar to a usual baryon/meson ratio,
in order to investigate whether there exists enhanced production
for a four-quark hadron compared to a normal hadron at
intermediate transverse momentum region due to some reasons, e.g.,
hadron production mechanisms or transverse momentum distributions
of constituent quarks. We have further evaluated transverse
momentum distribution ratios between multi-charmed hadrons and a
singly charmed baryon, the $\Lambda_c$. Lastly, we have discussed
the transverse momentum distribution of the $X(3872)$ in a
four-quark state, the $X_4$, and that of the $X(3872)$ in a
two-quark state, the $X_2$.

We find that yields decrease with increasing number of charm and
light quarks in multi-charmed hadrons in both the statistical and
coalescence models as expected. However, when the $X(3872)$ is
considered to be a normal meson composed of a charm and and an
anti-charm quark, the $X_2$, the yield in the coalescence model is
almost same as that in the statistical model. It is interesting to
notice that the yield of the $X_4$ is comparable to that of the
$\Omega_{scc}$ in the coalescence model when the feed-down
contribution of the $\Omega_{scc}^*$ to the $\Omega_{scc}$ is not
taken into account; the effect of including two more light quarks
of the constituent mass 350 MeV is comparable to that of adding
one more strange quark of the mass 500 MeV.

The yield of multi-charmed hadrons in the quark coalescence model
is found to be smaller than that in the statistical model,
reflecting the suppression effects in the quark coalescence
process. Among the yield ratio between charmed hadrons we find
that yield ratios involving the $\Omega_{ccc}$, or
$\Omega_{ccc}/\Xi_{cc}$, $\Omega_{ccc}/\Omega_{scc}$ and
$X_4/\Omega_{ccc}$ at LHC are always larger than those at RHIC in
both the statistical and coalescence model. For other ratios
without the $\Omega_{ccc}$ ratios at RHIC are comparable to or
larger than those at LHC.

Transverse momentum distributions of charmed hadrons at LHC are
found to be larger than those at RHIC due to the larger number of
charm and light quarks available at LHC compared to that at RHIC.
We find that the transverse momentum distribution of the $X_4$
meson is larger than that of the $\Omega_{ccc}$ baryon by two
orders of magnitude at both RHIC and LHC. The effect from the much
smaller abundance of charm quarks in the system compared to that
of light quarks by a factor of hundreds at RHIC and LHC overwhelms
the typical larger probability for forming a hadron composed of
three charm quarks compared to the relatively smaller possibility
for forming a hadron with four quarks, leading to the smaller
transverse momentum distribution of the $\Omega_{ccc}$ baryon
compared to that of the $X(3872)$ meson.

We note that some transverse momentum distribution ratios between
multi-charmed hadrons have a peak in the intermediate transverse
momentum region, similar to that in the transverse momentum
distribution ratio between an anti-proton and a pion. We find that
the peak appearing in the transverse momentum distribution ratio
between multi-charmed baryons is related to a kind of quark
constituents participating in hadron production; specific
combinations between light quarks in thermal equilibrium with an
exponential transverse momentum distribution and charm quarks not
in thermal equilibrium in a system with their transverse momentum
distributions of both a power law type and an exponential. We also
notice that the position of the peak is located at higher
transverse momentum when more heavier quarks are involved; the
peak in the ratio $ccc/ccs$ is shifted to the higher transverse
momentum compared to that in the ratio, $ccc/ccq$. The peak
located at the higher transverse momentum for hadrons with heavier
quarks confirms the argument that the momentum of heavy quark
hadrons is mostly carried by heavy quarks due to their heavier
mass. In addition we further observe that the location of the peak
is closely related to the transverse momentum distribution of the
spectator quark in the ratio.

The transverse momentum distribution ratio between the $X(3872)$
and the $\Xi_{cc}$ is noticeable since it has a peak in the
intermediate transverse momentum region, leaving light quarks in
the ratio, $q\bar{q}/q$ similar to the ratio between anti-protons
and pions, or that between the $\Lambda_c$ and the $D^0$. We note
that the ratio increases up to about 0.20 at both RHIC and LHC, no
significant enhanced production of the heavier $X(3872)$ meson
compared to the $\Xi_{cc}$ baryon in the intermediate transverse
momentum region.

The transverse momentum distribution ratio between the
multi-charmed hadron and the $\Lambda_c$ is found to be much
smaller than unity, reflecting the small possibility to coalesce
more charm quarks to form a multi-charmed hadron, but to have a
peak in all the ratios. We argue that a peak appears in the ratio
involving charm and light quarks, especially when a remaining
charm quark is in the numerator and a light quark remains in the
denominator after cancelling common quarks, e.g., $c/q$. We also
note that the ratio of the $\Omega_{ccc}$ with the doubly-charmed
baryon has the broader peak at the higher transverse momentum than
that of the $\Omega_{ccc}$ with the $\Lambda_c$.

The ratio between the $X(3872)$ in a two-quark state, the $X_2$
and charmed baryons are calculated to be completely different from
that between the $X(3872)$ in a four-quark state, the $X_4$ and
charmed baryons. We consider that we can infer the quark content
of the $X(3872)$ by measuring the transverse momentum distribution
ratio between the $X(3872)$ and various charmed baryons. It is
interesting to observe the peak in the ratio between the
$\Omega_{ccc}$ and the $X_2$ similar to that in the ratio between
the anti-proton and the pion, attributable to similar quark
contents in two ratios, $\bar{q}\bar{q}/q$ for the $\bar{p}/\pi$
and $cc/\bar{c}$ for the $\Omega_{ccc}/X_2$. We hope that we
compare the transverse momentum distribution of the $X(3872)$
yield obtained here to that measured in relativistic heavy ion
collision experiments as well as that obtained in the statistical
hadronization model in the near future, and that we identify the
quark structure of the $X(3872)$ meson.

In summary we find that yields of multi-charmed hadrons in heavy
ion collisions at RHIC and LHC are large enough, and thereby not
only multi-charmed hadrons observed so far, e.g., the $\Xi_{cc}$
but also those which have not been observed yet, are expected to
be discovered more easily in heavy ion collisions. On the other
hand the transverse momentum distribution of multi-charmed hadrons
is found to keep the information of their constituent quarks at
the moment of hadron production very well, and the effects of
constituent quarks, especially charm quarks on the transverse
momentum distribution of the multi-charmed hadron become more
visible in the transverse momentum ratio between various
multi-charmed hadrons.

Charm quarks carrying most of the total momentum of charmed
hadrons due to their heavier mass compared to that of light
quarks, determine both the position and broadness of the peak in
the transverse momentum distribution ratio between charmed quark
hadrons. The transverse momentum distribution ratio reflects the
interplay between quark contents of two corresponding hadrons, and
the peak in the transverse momentum distribution ratio between
multi-charmed hadrons appears only under certain circumstances,
presenting us with meaningful information on constituent quarks.
Therefore we expect that studying both transverse momentum
distributions of multi-charmed hadrons themselves and transverse
momentum distribution ratios between various multi-charmed hadrons
provide us useful information on hadron production mechanism
involving charm quarks in heavy ion collisions.

\section*{Acknowledgements}

S. Cho was supported by the National Research Foundation of Korea
(NRF) grant funded by the Korea government (MSIP) (No.
2016R1C1B1016270) and supported by the National Research
Foundation of Korea (NRF) grant funded by the Korea government
(MSIT) (No. 2018R1A5A1025563). S. H. Lee was supported by Samsung
Science and Technology Foundation under Project Number
SSTF-BA1901-04.

\appendix

\section{Relative coordinates for different internal structures of
the X(3872) meson}

 When the $X(3872)$ meson is in a tetra-quark state, there exist
various possibilities to define relative coordinates, and
therefore relative momenta in explaining the internal structure of
the $X(3872)$ meson \cite{Brink:1994ic}. As introduced in Sec.
III, the most common space coordinates used to describe four-quark
states are,
\begin{eqnarray}
&& \vec R=\vec r_l'+\vec r_{\bar{l}}'+\vec r_c'+\vec
r_{\bar{c}}', \nonumber \\
&& \vec r_1'=\vec r_l'-\vec r_{\bar{l}}', \nonumber \\
&& \vec r_2'=\frac{m_l\vec r_l'+m_{\bar{l}'}\vec
r_{\bar{l}}'}{m_l+m_{\bar{l}}}-\vec r_c, \nonumber \\
&& \vec r_3'=\frac{m_l\vec r_l'+m_{\bar{l}}\vec
r_{\bar{l}}'+m_c\vec r_c'}{m_l+m_{\bar{l}}+m_c}-\vec r_{\bar{c}}',
\label{coord1}
\end{eqnarray}
and the corresponding relative transverse momenta are,
\begin{eqnarray}
&& \vec k=\vec p_{lT}'+\vec p_{\bar{l}T}'+\vec p_{cT}'+\vec
p_{\bar{c}T}', \nonumber \\
&& \vec k_1=\frac{m_{\bar{l}}\vec p_{lT}'-m_l\vec
p_{\bar{l}T}'}{m_l+m_{\bar{l}}}, \nonumber \\
&& \vec k_2=\frac{m_c(\vec p_{lT}'+\vec
p_{\bar{l}T}')-(m_l+m_{\bar{l}})
\vec p_{cT}'}{m_l+m_{\bar{l}}+m_c}, \nonumber \\
&& \vec k_3=\frac{m_{\bar{c}}(\vec p_{lT}'+\vec p_{\bar{l}T}'+\vec
p_{cT}')-(m_l+m_{\bar{l}}+m_c)\vec
p_{\bar{c}T}'}{m_l+m_{\bar{l}}+m_c+m_{\bar{c}}}, \label{moment1}
\end{eqnarray}
with reduced masses for the above configurations,
\begin{eqnarray}
&& \mu_1=\frac{m_lm_{\bar{l}}}{m_l+m_{\bar{l}}}, \quad \mu_2=
\frac{(m_l+m_{\bar{l}})m_c}{m_l+m_{\bar{l}}+m_c}, \nonumber \\
&& \mu_3=\frac{(m_l+m_{\bar{l}}+m_c)m_{\bar{c}}}{m_l+m_{\bar{l}}
+m_c+m_{\bar{c}}},
\end{eqnarray}
which describe relative coordinates of quarks in the $X(3872)$
meson; the distance between any two quarks, $\vec r_1'$, that
between the center of mass for previously chosen two quarks and
the third quark, $\vec r_2'$, and that between the center of mass
for three quarks and the remaining quark fourth quark, $\vec
r_3'$. Here two light quarks are chosen for the first relative
coordinate $\vec r_1'$ in Eq. (\ref{coord1}) but it does not
matter whether other combination of quarks, e.g., two charm quarks
or one light and one charm quarks, is chosen.

\begin{figure}[!t]
\begin{center}
\subfigure[]{
\includegraphics[width=0.12\textwidth]{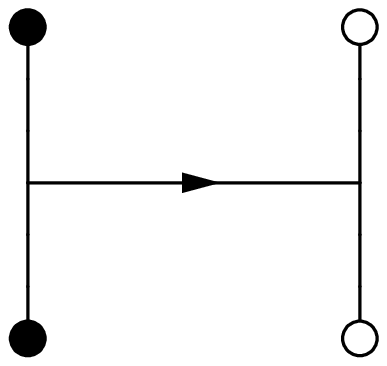}}
\quad \subfigure[]{
\includegraphics[width=0.11\textwidth]{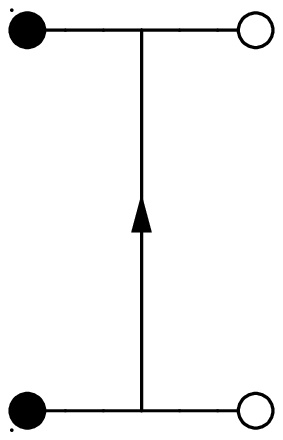}}
\quad \subfigure[]{
\includegraphics[width=0.12\textwidth]{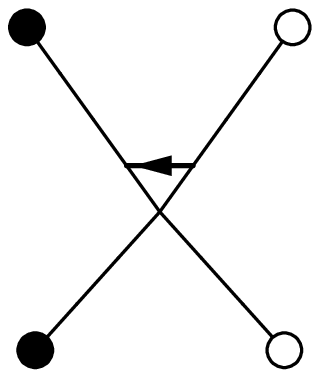}}
\end{center}
\caption{Alternative relative coordinates for a four-quark
$X(3872)$ meson. Filled circles represent quarks and empty circles
represent anti-quarks. } \label{relative_coordinates}
\end{figure}

On the other hand, if we want to explain the $X(3872)$ meson
formed from the coalescence of two quark pairs, especially when we
want to describe the most probable configuration of the $X(3872)$
meson which easily decays strongly to $J/\Psi$ and $\rho$ mesons,
we find that the following relative coordinates and momenta would
be more appropriate,
\begin{eqnarray}
&& \vec R=\vec r_l'+\vec r_{\bar{l}}'+\vec r_c'+\vec
r_{\bar{c}}', \nonumber \\
&& \vec r_1'=\vec r_l'-\vec r_{\bar{l}}', \nonumber \\
&& \vec r_2'=\vec r_c'-\vec r_{\bar{c}}', \nonumber \\
&& \vec r_3'=\frac{m_l\vec r_l'+m_{\bar{l}}\vec r_{\bar{l}}'}
{m_l+m_{\bar{l}}}-\frac{m_c\vec r_c'+m_{\bar{c}}\vec r_{\bar{c}}}
{m_c+m_{\bar{c}}}, \label{coord2}
\end{eqnarray}
and
\begin{eqnarray}
&& \vec k=\vec p_{lT}'+\vec p_{\bar{l}T}'+\vec p_{cT}'+\vec
p_{\bar{c}T}', \nonumber \\
&& \vec k_1=\frac{m_{\bar{l}}\vec p_{lT}'-m_l\vec
p_{\bar{l}T}'}{m_l+m_{\bar{l}}}, \nonumber \\
&& \vec k_2=\frac{m_{\bar{c}}\vec p_{cT}'-m_c\vec
p_{\bar{c}T}'}{m_c+m_{\bar{c}}}, \nonumber \\
&& \vec k_3=\frac{(m_c+m_{\bar{c}})(\vec p_{lT}'+\vec
p_{\bar{l}T}')-(m_l+m_{\bar{l}})(\vec p_{cT}'+\vec
p_{\bar{c}T}')}{m_l+m_{\bar{l}}+m_c+m_{\bar{c}}}, \nonumber \\
\label{moment2}
\end{eqnarray}
with reduced masses for the above configurations,
\begin{eqnarray}
&& \mu_1=\frac{m_lm_{\bar{l}}}{m_l+m_{\bar{l}}}, \quad \mu_2=
\frac{m_c m_{\bar{c}}}{m_c+m_{\bar{c}}}, \nonumber \\
&& \mu_3=\frac{(m_l+m_{\bar{l}})(m_c+m_{\bar{c}})}{m_l+m_{\bar{l}}
+m_c+ m_{\bar{c}}}.
\end{eqnarray}
In the coordinate, Eq. (\ref{coord2}), two pairs of two quarks are
chosen first, and then the distance between the center of mass
between two pairs of two quarks is taken. We show a diagram
describing a system of relative coordinates, Eq. (\ref{coord2}) in
Fig. \ref{relative_coordinates} (b) similar to the figure as shown
in Ref. \cite{Brink:1994ic}. Filled circles represent quarks, and
empty circles represent anti-quarks, and Fig.
\ref{relative_coordinates} (b) and (c) correspond to meson-meson
channels, a direct and an exchange channel, respectively. If the
upper circles represent heavy quarks and lower circles represent
the light quarks, the system of coordinates, Eq. (\ref{coord2})
corresponds to the direct channel good for describing the
formation of the $X(3872)$ meson from the $J/\Psi$ and $\rho$
meson whereas the following exchange channel would be good for
describing the decay of the $X(3872)$ meson to $D^*$ and $\bar{D}$
mesons or $\bar{D}^*$ and $D$ mesons,
\begin{eqnarray}
&& \vec R=\vec r_l'+\vec r_{\bar{l}}'+\vec r_c'+\vec
r_{\bar{c}}', \nonumber \\
&& \vec r_1'=\vec r_c'-\vec r_{\bar{l}}', \nonumber \\
&& \vec r_2'=\vec r_l'-\vec r_{\bar{c}}', \nonumber \\
&& \vec r_3'=\frac{m_c\vec r_c'+m_{\bar{l}}\vec r_{\bar{l}}'}
{m_c+m_{\bar{l}}}-\frac{m_l\vec r_l'+m_{\bar{c}}\vec r_{\bar{c}}}
{m_l+m_{\bar{c}}}, \label{coord3}
\end{eqnarray}
and
\begin{eqnarray}
&& \vec k=\vec p_{lT}'+\vec p_{\bar{l}T}'+\vec p_{cT}'+\vec
p_{\bar{c}T}', \nonumber \\
&& \vec k_1=\frac{m_{\bar{l}}\vec p_{cT}'-m_c\vec
p_{\bar{l}T}'}{m_c+m_{\bar{l}}}, \nonumber \\
&& \vec k_2=\frac{m_{\bar{c}}\vec p_{lT}'-m_l\vec
p_{\bar{c}T}'}{m_l+m_{\bar{c}}}, \nonumber \\
&& \vec k_3=\frac{(m_c+m_{\bar{l}})(\vec p_{lT}'+\vec
p_{\bar{c}T}')-(m_l+m_{\bar{c}})(\vec p_{cT}'+\vec
p_{\bar{l}T}')}{m_l+m_{\bar{l}}+m_c+m_{\bar{c}}}, \nonumber \\
\label{moment3}
\end{eqnarray}
with reduced masses for this configuration,
\begin{eqnarray}
&& \mu_1=\frac{m_cm_{\bar{l}}}{m_c+m_{\bar{l}}}, \quad \mu_2=
\frac{m_l m_{\bar{c}}}{m_l+m_{\bar{c}}}, \nonumber \\
&& \mu_3=\frac{(m_c+m_{\bar{l}})(m_l+m_{\bar{c}})}{m_l+m_{\bar{l}}
+m_c+ m_{\bar{c}}}.
\end{eqnarray}
Similarly, a system of coordinates for Fig.
\ref{relative_coordinates} (a) becomes,
\begin{eqnarray}
&& \vec R=\vec r_l'+\vec r_{\bar{l}}'+\vec r_c'+\vec
r_{\bar{c}}', \nonumber \\
&& \vec r_1'=\vec r_c'-\vec r_l', \nonumber \\
&& \vec r_2'=\vec r_{\bar{l}}'-\vec r_{\bar{c}}', \nonumber \\
&& \vec r_3'=\frac{m_c\vec r_c'+m_l\vec r_l'}
{m_c+m_l}-\frac{m_{\bar{l}}\vec r_{\bar{l}}'+m_{\bar{c}}\vec
r_{\bar{c}}'}{m_{\bar{l}}+m_{\bar{c}}}, \label{coord4}
\end{eqnarray}
and
\begin{eqnarray}
&& \vec k=\vec p_{lT}'+\vec p_{\bar{l}T}'+\vec p_{cT}'+\vec
p_{\bar{c}T}', \nonumber \\
&& \vec k_1=\frac{m_l\vec p_{cT}'-m_c\vec
p_l'}{m_c+m_l}, \nonumber \\
&& \vec k_2=\frac{m_{\bar{c}}\vec p_{\bar{l}T}'-m_{\bar{l}}\vec
p_{\bar{c}T}'}{m_{\bar{l}}+m_{\bar{c}}}, \nonumber \\
&& \vec k_3=\frac{(m_c+m_l)(\vec p_{\bar{l}T}'+\vec
p_{\bar{c}T}')-(m_{\bar{l}}+m_{\bar{c}})(\vec p_{cT}'+\vec
p_{lT}')}{m_l+m_{\bar{l}}+m_c+m_{\bar{c}}}, \nonumber \\
\label{moment4}
\end{eqnarray}
with reduced masses,
\begin{eqnarray}
&& \mu_1=\frac{m_cm_l}{m_c+m_l}, \quad \mu_2=
\frac{m_{\bar{l}}m_{\bar{c}}}{m_{\bar{l}}+m_{\bar{c}}}, \nonumber \\
&& \mu_3=\frac{(m_c+m_l)(m_{\bar{l}}+m_{\bar{c}})}{m_l+m_{\bar{l}}
+m_c+ m_{\bar{c}}}.
\end{eqnarray}

As has been discussed in Refs. \cite{Cho:2010db, Cho:2011ew,
Cho:2017dcy} it has been found that the yield of an exotic hadron
depends on the internal structure of the exotic hadron. We
therefore consider that a transverse momentum distribution of an
exotic hadron should be somehow dependent on its internal
structure. Moreover, we have alternative relative coordinates in
describing a four-quark hadron, and therefore we need to check
whether the transverse momentum distribution of an exotic hadron,
here the $X(3872)$ meson would be dependent also on each relative
coordinate.

When we look at the equation for the transverse momentum
distribution of the $X(3872)$ meson, Eq. (\ref{4CoalTrans}), we
see that the only part where the relative coordinates play an
important role is the Wigner function. As has been pointed out in
Ref. \cite{Cho:2014xha} different wave functions contribute to the
transverse momentum distribution differently through the Wigner
function for hadrons composed of the same kind and number of
constituents. If we just consider here the Gaussian Wigner
function in a $s$-wave applied in the analysis, Eq.
(\ref{4CoalTrans}), we find that the relative momentum part
becomes,
\begin{equation}
e^{-\sigma_1^2k_1^2-\sigma_2^2k_2^2-\sigma_3^2k_3^2}=e^{-\frac{1}
{\omega_c}\Big(\frac{k_1^2}{\mu_1}+\frac{k_2^2}{\mu_2}
+\frac{k_3^2}{\mu_3} \Big)}.
\end{equation}
with the relation, $\sigma^2=1/\mu\omega$.  When we put relative
momenta corresponding to alternative relative coordinates, Eqs.
(\ref{coord1}), (\ref{coord2}), (\ref{coord3}), and (\ref{coord4})
with the reduced mass also corresponding to each relative
coordinate, we obtain,
{\allowdisplaybreaks
\begin{eqnarray}
&& -\sigma_1^2k_1^2-\sigma_2^2k_2^2-\sigma_3^2k_3^2=
-\frac{1}{\omega_c}\bigg(\frac{k_1^2}{\mu_1}+\frac{k_2^2}{\mu_2}
+\frac{k_3^2}{\mu_3} \bigg) \nonumber \\
&& =-\frac{1}{\omega_c} \frac{1}{m_l+m_{\bar{l}}+m_c+m_{\bar{c}}}
\bigg(\frac{m_{\bar{l}}+m_c+m_{\bar{c}}}{m_l}{p_{lT}'}^2 \nonumber \\
&& +\frac{m_c+m_{\bar{c}}+m_l}{m_{\bar{l}}}{p_{\bar{l}T}'}^2+
\frac{m_{\bar{c}}+m_l+m_{\bar{l}}}{m_c}{p_{cT}'}^2 \nonumber \\
&& + \frac{m_l+m_{\bar{l}}+m_c}{m_{\bar{c}}}{p_{\bar{c}T}'}^2
-2(\vec p_{lT}'\cdot\vec p_{\bar{l}T}'+\vec p_{\bar{l}T}'\cdot\vec
p_{cT}' \nonumber \\
&& +\vec p_{cT}'\cdot\vec p_{\bar{c}T}'+\vec p_{lT}'\cdot\vec
p_{cT}' +\vec p_{\bar{l}T}'\cdot\vec p_{cT}'+\vec
p_{\bar{l}T}'\cdot\vec p_{\bar{c}T}')\bigg), \label{exponents}
\end{eqnarray} }
the same result regardless of any relative coordinates; the
argument in Eq. (\ref{exponents}) is symmetric between four
quarks, and thus does not change under any exchange between
constituent quarks in the $X(3872)$ meson. Therefore, we find that
the transverse momentum distribution for a four-quark state does
not depend on the choice of relative coordinates. In other words,
we cannot identify the internal structure of a hadron as shown in
Fig. \ref{relative_coordinates} based on transverse momentum
distributions of the hadron. We, however, still see that if the
four-quark state has the internal relative momentum, e.g.,
$p$-wave or $d$-wave, and so on, then the transverse momentum
distribution would be dependent on it since there exists an
additional term outside the argument in the exponential Wigner
function as shown in Eq. (\ref{CoalTransX2}), and thereby the
yield is expected to be affected by the relative internal momentum
as discussed in Refs. \cite{Cho:2010db, Cho:2011ew, Cho:2017dcy}.

We also show the argument similar to that shown in Eq.
(\ref{exponents}) for three-quark baryons,

\begin{eqnarray}
&& -\sigma_1^2k_1^2-\sigma_2^2k_2^2=
-\frac{1}{\omega}\bigg(\frac{k_1^2}{\mu_1}+\frac{k_2^2}{\mu_2}
\bigg) \nonumber \\
&& =-\frac{1}{\omega} \frac{1}{m_1+m_2+m_3}
\bigg(\frac{m_1+m_2}{m_3}{p_{3T}'}^2 \nonumber \\
&& +\frac{m_2+m_3}{m_1}{p_{1T}'}^2+\frac{m_3+m_1}{m_2}{p_{2T}'}^2
\nonumber \\
&& -2(\vec p_{1T}'\cdot\vec p_{2T}'+\vec p_{2T}'\cdot\vec p_{3T}'
+\vec p_{3T}'\cdot\vec p_{1T}')\bigg), \label{exponents3}
\end{eqnarray}
which is symmetric between three quarks, and thus does not change
under any exchange between constituent quarks in the three-quark
baryon. We also see that if the three-quark baryon has the
internal relative momentum, then the transverse momentum
distribution would be dependent on it, and thereby the yield is
expected to be affected by the relative internal momentum.

\end{document}